\documentclass[3p,final]{elsarticle}

\usepackage{amssymb}
\usepackage{amsmath}

\usepackage[figuresright]{rotating}

\newcommand{\ac}{$\alpha+^{12}${\rm C}}
\newcommand{\apb}{$\alpha+^{208}${\rm Pb}}
\newcommand{\oca}{$^{16}{\mathrm O}+ ^{44}${\rm Ca}}
\newcommand{\beq}{\begin{eqnarray}}
\newcommand{\eeq}{\end{eqnarray}}
\newcommand{\ecm}{\mbox{$E_{\mathrm c.m.}$}}

\begin{document}

\begin{frontmatter}




\title{An $R$-matrix package for coupled-channel problems in nuclear physics}

\author[label1]{P. Descouvemont\fnref{label2}}
\ead{pdesc@ulb.ac.be}
\address[label1]{Physique Nucl\'{e}aire Th\'{e}orique et Physique Math\'{e}matique, CP229\\
	Universit\'{e} Libre de Bruxelles (ULB), B1050 Brussels, Belgium}
\fntext[label2]{Directeur de Recherches FNRS}

\begin{abstract}
We present an $R$-matrix Fortran package to solve coupled-channel problems in nuclear physics.
The basis functions are chosen as Lagrange functions, which permits simple calculations of	the matrix elements. The main input are the coupling potentials at some nucleus-nucleus distances, specified
by the program. The program provides the collision matrix and, optionally, the associated wave function.
The present method deals with open and closed channels simultaneously, without numerical
instability associated with closed channels. It can also solve coupled-channel problems for non-local	potentials. Long-range potentials can be treated with propagation techniques, which
significantly speed up the calculations. We first present an overview of the $R$-matrix theory, and of the Lagrange-mesh method. A description of the package and its installation on a UNIX machine is then provided.
Finally, five typical examples are discussed.
\end{abstract}




\end{frontmatter}
\vspace{1 cm}

{\bf PROGRAM SUMMARY}

\begin{small}
\noindent
{\em Manuscript Title:} An $R$-matrix package for coupled-channel problems in nuclear physics                                       \\
{\em Author:} P. Descouvemont                                                \\
{\em Program Title:} rmatrix                                          \\
{\em Journal Reference:}                                      \\
{\em Catalogue identifier:}                                   \\
{\em Licensing provisions:} none                                   \\
{\em Programming language:} Fortran 90                                   \\
{\em RAM:} memory usage strongly depends on the nature of the problem (number of channels, size of the basis).                         \\
{\em Keywords:} scattering theory, $R$-matrix method, Lagrange meshes, coupled-channel problems.  \\
{\em Classification:}  17.8	Nuclear Reaction - General                                       \\
{\em External routines/libraries:} if available, the LAPACK library can be used  for matrix inversion (faster than
the subroutine included in the package).                          \\
{\em Nature of problem:} solving coupled-channel problems for positive energies; the package provides the collision matrix
and the associated wave function for given energy, spin and parity. \\
   \\
{\em Solution method:} the coupled-channel system is solved with the $R$-matrix method, and  Lagrange functions are adopted.
Propagation of the $R$-matrix can be performed.\\
{\em Restrictions:} for non-local potentials, the propagation technique cannot be used.\\
{\em Running time:} strongly depends on the nature of the problem. For single-channel calculations,
	the running type is typically less than 1 s. For large scale calculations (typically more than 100 channels), the
	running time may increase up to several minutes on a multi-CPU Linux machine.\\
\end{small}

\section{Introduction}
\label{intro}

Many problems in quantum physics require the solution of a second-order differential system.  
Nuclear \cite{Ta65,Sa83,Th88} and atomic \cite{MTG98,PN99,Bu11} collisions are obvious examples 
(see also Ref. \cite{CH13}).  Solving a coupled-channel system is also necessary in other 
applications, such as the description of three-body systems \cite{ZDF93,TND04}.  The Resonating Group 
Method (RGM, see Ref.\ \cite{Ho77}) is used in nuclear physics to describe nucleus-nucleus collisions from
$A$-body wave functions, and takes antisymmetrization exactly into account. 
It provides a system involving non-local potentials \cite{Ho77,TMO07,QN08}.  

Solving a coupled-channel system for bound states, i.\ e.\ for energies below all threshold energies, is 
relatively simple, and can be addressed by different methods (see, for example, Ref.\ \cite{KNG01}).  In particular, 
variational methods using large bases \cite{HKK03,VKR06,MBH13} provide 
accurate solutions for bound-state problems.  Various improvements, such as a random selection of the basis functions \cite{VS95},
have been proposed to optimize the choice of basis states. The main difficulty in bound-state calculations is to deal with the
diagonalization of extremely large bases.

Scattering states, however, are more complicated owing to the long range behavior of the wave functions.  
Finite-difference methods, based on the Numerov algorithm \cite{Ra72,BP82b,Th87}, can be used to integrate the system until large distances, where the wave function
reaches its asymptotic behaviour. The matching to Coulomb functions then provides the collision matrix, which is subsequently
used to compute various cross sections. In contrast, the $R$-matrix method is a well-established tool to complement variational calculations,
frequently used to define bound states, to scattering states.
However, as variational calculations involve square integrable functions, any finite combination tends to zero at large distances. 
 
The original idea of the $R$-matrix method is due to Wigner \cite{Wi46} with the aim of describing resonances.  The $R$-matrix theory describes scattering states between interacting particles, which can be nuclei, atoms, molecules.  It is based on the division of the configuration space in two regions.  In the internal region, both particles interact through some potential, which depends on the model.  In the external region, possible antisymmetrization effects and non-monopole Coulomb terms are neglected.  The border is called the channel radius and must be chosen large enough to comply with these requirements.  Extensions to the scattering of three particles have been developed \cite{TDE00,DTB06}, but will not be presented here. Three-body continuum wave functions are required, for example, in
breakup calculations \cite{BCD09}.

The $R$-matrix method has been presented in many reviews or books.  It has been developed both in 
atomic \cite{Bu11} and nuclear \cite{LT58,DB10} physics.  As basis functions, we use here Lagrange functions \cite{Ba15} which permit fast calculations of the matrix elements \cite{HSV98,DB10}.  Another important advantage is that non-local potentials are easily implemented \cite{HRB02}. In addition, closed channels are treated in a straightforward way, in contrast with finite-difference methods,
where the exponentially growing component raises strong numerical problems.
Our main goal here is to provide 
the user with an efficient routine, which can be easily implemented in any code, as soon as 
the coupling potentials are known.  Previous codes are essentially focused on atomic physics or only consider propagation of the $R$-matrix across several intervals \cite{BBM82,Mo84,MTG98}.

Notice that the $R$-matrix theory can be used in another variant, known as the {\em phenomenological} 
$R$-matrix \cite{LT58,BHT68,DB10}.  Although the origins of both variants are common, the goal 
of the phenomenological approach is to fit the $R$-matrix parameters to experimental data, such as elastic-scattering or radiative-capture cross sections.  This variant is essentially used 
to determine resonance parameters at low energies and, in nuclear astrophysics, to extrapolate 
cross sections down to stellar energies, where, in general, no data exist.

We present first a brief overview of the $R$-matrix theory in Section \ref{sec_rmat}.  Sections 
\ref{manual} and \ref{sec_list} provide a description of the package. We give typical examples in Section 
\ref{sec_exam}, with the outputs presented in Appendices.  Concluding remarks are presented in Section \ref{sec_conc}.

\section{Brief overview of the $R$-matrix method}
\label{sec_rmat}

\subsection{The collision matrix}

We present here a short description of the theory, and we refer the reader to 
Refs.\ \cite{Bu11,DB10} for more detail.  Let us consider a coupled-channel system 
for two particles
\begin{eqnarray}
\sum_{j=1}^{N_c}\bigl(T_i \delta_{ij}+V_{ij}(r)\bigr)u_j(r)=(E-E_i)u_i(r),
\label{eq1}
\end{eqnarray}
where $r$ is the relative coordinate, $N_c$ is the number of channels, $E$ is 
the scattering energy, and $E_i$ is the threshold energy of channel $i$.  
Notice that we define a ``channel" as characterized by a threshold energy $E_i$ and by an orbital angular
momentum $L_i$. This definition is well adapted to numerical aspects. However, it should not be confused with
``physical channels", characterized by the quantum numbers (energy, spin and parity) of the colliding nuclei.
We assume that all energies $(E,E_i)$ are given with respect to $E_{\omega}$, the energy of the entrance channel
$\omega$.

In (\ref{eq1}), $V_{ij}(r)$ are the coupling potentials (symmetric), which may be real or complex. 
The kinetic-energy operator reads 
\begin{eqnarray}
T_i=-\frac{\hbar^2}{2\mu}\biggl( \frac{d^2}{dr^2}-\frac{L_i(L_i+1)}{r^2}\biggr),
\label{eq2}
\end{eqnarray}
where $\mu$ is the reduced mass (we assume that it does not depend on the channel), 
and $L_i$ the orbital momentum in channel $i$.  For the sake of clarity we 
consider now local potentials only, but a generalization to non-local potentials is possible, and
will be presenter later.  

As mentioned in the introduction, the coupling potentials may originate from various models. 
However, the $R$-matrix method is a tool to solve (\ref{eq1}) for the continuum, and does not depend on
the physics of the problem. The main requirement is that,
at large distances, the potentials tend to
\begin{eqnarray}
V_{ij}(r)\rightarrow\frac{Z_1 Z_2 e^2}{r}\delta_{ij},
\label{eq3}
\end{eqnarray}
where $Z_1 e$ and $Z_2 e$ are the charges of the colliding nuclei.  This assumption implicitly
defines the
lower limit for the choice of the channel radius.
  
For a bound state, all channels are closed $(E<E_i)$, and long-range contributions 
to the wave function are, in general, neglected.
A scattering state (including resonances) is characterized by at least one 
open channel.  At large distances, i.\ e.\ when Eq.\ (\ref{eq3}) is valid, the solutions 
of the coupled-channel system (\ref{eq1}) are given by
\begin{eqnarray}
u_{i,\rm ext}^{\omega}(r)=
\left\{\begin{array}{ll}
	v_{i}^{-1/2} \Bigl( I_{L_i} (k_{i} r)\delta_{i \omega} -  O_{L_i} (k_{i} r)U_{i \omega} \Bigr)
	& {\rm \ for\ } E> E_i, \\
	A_{i}^{\omega} W_{-\eta_{i},L_i+1/2}(2k_{i}r) & {\rm \ for\ } E<E_i .
\end{array} \right .
\label{eq4}
\end{eqnarray}
In these definitions, $v_i$ and $k_i$ are the velocity and wave number in channel 
$i$  $(k_i^2=2\mu\vert E-E_i \vert /\hbar^2)$, and $\omega$ is the entrance channel.  
Functions $I_{L_i}(x)$ and $O_{L_i}(x)$ are the incoming and outgoing Coulomb functions \cite{Th10}, 
and $W_{a,b}(x)$ is the Whittaker function \cite{Ol10}.  Equations (\ref{eq4}) define the collision 
matrix $\pmb{U}$, which is used to compute cross sections \cite{Th88,CH13}.  
The collision matrix is symmetric, and unitary for real potentials.  The other 
output from the calculations are the amplitudes of closed channels $A_{i}^{\omega}$. 
Although some solutions have been proposed in the literature (see e.g. Ref.\ \cite{BP82b}), 
including closed channels in finite-difference
methods raises numerical problems, due to the exponential behaviour of the associated wave-function
components. At energies close to breakup channels, however, closed channels may be expected to play a significant
role. In the $R$-matrix theory, open and closed channels are treated on an equal footing.
 
Notice that, in the literature, matrix $\pmb{U}$ is also called the scattering matrix.
We follow here the naming used by Lane and Thomas \cite{LT58}. Another important comment
deals with the normalization (\ref{eq4}). This choice ensures that the collision matrix is 
symmetric. An alternative is to define the asymptotic part of open-channel components as
\begin{eqnarray}
u_{i,\rm ext}^{\omega}(r)=v_{\omega}^{-1/2} \biggl(
 I_{L_i} (k_{i} r)\delta_{i \omega} -  O_{L_i} (k_{i} r)\tilde{U}_{i \omega}
 \biggr),
\label{eq4b}
\end{eqnarray}
with 
\begin{eqnarray}
\tilde{U}_{i j}=\biggl( \frac{k_j}{k_i} \biggr) ^{1/2}U_{ij}.
\label{eq4c}
\end{eqnarray} 
This choice is adopted, for example, in Ref.\ \cite{Th88}. In that case, matrix $\tilde{\pmb{U}}$ 
is not symmetric.

The basic idea of the $R$-matrix method is to solve system (\ref{eq3}) on a limited 
interval, defined by the channel radius $a$.  In the internal region $(r\leq a) $, 
the wave functions are expanded over $N$ basis states $\varphi_n(r)$ as
\begin{eqnarray}
u_{i,\rm int}^{\omega}(r)=\sum_{n=1}^N f^{\omega}_{in}\, \varphi_n(r).
\label{eq5}
\end{eqnarray}
These basis functions can be orthogonal or not.
The unknown quantities of a scattering problem are therefore the coefficients $f^{\omega}_{in}$, the collision 
matrix $U_{ij}$, and the closed-channel amplitudes $A_{i}^{\omega}$.  These quantities 
are obtained from the $R$-matrix theory as described below.

The kinetic-energy operator is not hermitian over a finite range $[0,a]$.  
This issue is solved by introducing the Bloch operator \cite{Bl57}
\begin{eqnarray}
{\cal L}(B_i)=\frac{\hbar^2}{2\mu}\delta(r-a)\biggl(\frac{d}{dr}-\frac{B_i}{r}\biggr),
\label{eq6}
\end{eqnarray}
which is also called surface operator since it acts at $r=a$ only.
Constants $B_i$ are boundary-condition parameters, and may depend on the channel.  For real $B_i$ values, 
the operator $T_i+{\cal L}(B_i)$ is hermitian over the interval $[0,a]$.
These parameters are chosen here as
\begin{eqnarray}
B_i=
\left\{\begin{array}{ll}
0 & {\rm \ for\ open\ channels},  \\
2k_i a \frac{W'(2k_i a)}{W(2k_i a)} & {\rm \ for\ closed \ channels},
\end{array} \right .
\label{eq6b}
\end{eqnarray}
where the prime denotes the derivative with respect to the argument $2k_i a$. 

In the $R$-matrix approach, system (\ref{eq1}) is therefore replaced by
the Bloch-Schr\"odinger equation
\begin{eqnarray}
\sum_{j}\biggl[\bigl(T_i+{\cal L}(B_i)+E_i-E\bigr) \delta_{ij}+V_{ij}(r)\biggr]
u^{\omega}_{j,\rm int}={\cal L}(B_i)u^{\omega}_{i,\rm int}={\cal L}(B_i)u^{\omega}_{i,\rm ext},
\label{eq7}
\end{eqnarray}
where we have used the continuity property
\begin{eqnarray}
u^{\omega}_{i,\rm int}(a)=u^{\omega}_{i,\rm ext}(a).
\label{eq8}
\end{eqnarray}
The choice of the boundary-condition parameters $B_i$ is such that the last term of
Eq.\ (\ref{eq7}) vanishes for closed channels.
Equation (\ref{eq7}) shows that the Bloch operator also guarantees the 
continuity of the derivative through the differential operator contained in (\ref{eq6}).
At the channel radius, we have, from (\ref{eq7}),
\begin{eqnarray}
u^{\omega \prime}_{i,\rm int}(a)=u^{\omega \prime}_{i,\rm ext}(a).
\label{eq9}
\end{eqnarray}
Notice that this equality holds for the exact solution only. If the variational basis (\ref{eq5})
is not well adapted, the derivative of the wave function is not continuous at $r=a$. This problem
is addressed in detail in Ref.\ \cite{DB10}.

Originally, the $R$-matrix method was developed without the Bloch operator, 
but imposing a constant boundary condition on the basis functions
\begin{eqnarray}
\varphi_n'(a)=B_i.
\label{eq10}
\end{eqnarray}
A typical choice was $\varphi_n'(a)=0$.
This leads to a  discontinuity in the derivative if a finite 
number of basis functions is used.  With the Bloch operator, imposing such 
a condition is not necessary, and accurate wave functions can be obtained 
without any matching problem \cite{BRT83,Bu67,DB10}.  

Let us define matrix $\pmb{C}$ as
\begin{eqnarray}
C_{in,jm}=\langle \varphi_n \vert
(T_i +{\cal L}(B_i)+E_i-E) \delta_{ij}+V_{ij} \vert \varphi_m \rangle_{\rm int},
\label{eq11}
\end{eqnarray}
where index ''int" in the Dirac notation refers to an integral over the $[0,a]$ interval.  
For traditional choices of basis functions, these matrix elements require 
numerical quadratures, which can be time-consuming for large systems.  
Lagrange functions, however, do not require any integral and form an orthogonal basis, as it will be shown 
in the next subsection. 

The $R$-matrix is then defined by
\begin{eqnarray}
R_{ij}=\frac{\hbar^2}{2\mu a}
\sum_{n,m} \varphi_n(a)\bigl(\pmb{C}^{-1}\bigr)_{in,jm}\varphi_m(a),
\label{eq12}
\end{eqnarray}
and involves the basis functions at the channel radius.  
Numerically, the inversion of matrix $\pmb{C}$ represents the longest part of 
the calculation, in particular when it is complex.  This may become a major issue for many-channel 
calculations, and when many basis states are required.  Propagation 
methods \cite{LW76} are aimed to address the problem by splitting the $[0,a]$ interval 
in smaller subintervals.  This will be discussed in Subsect.\ 2.4.

Using the continuity of the wave function (\ref{eq8}) provides the collision matrix as 
\begin{eqnarray}
\pmb{U}=\bigl( \pmb{Z}_O\bigr)^{-1}\pmb{Z}_I,
\label{eq13}
\end{eqnarray}
where an element of matrix $\pmb{Z}_O$ reads
\begin{eqnarray}
\bigl( \pmb{Z}_O\bigr)_{ij}=(k_j a)^{-1/2}\biggl(
O_{L_i}(k_ia)\delta_{ij}-k_ja R_{ij}O'_{L_j}(k_ja)\biggr).
\label{eq14}
\end{eqnarray}
A similar definition holds for matrix $\pmb{Z}_I$, with function $O_{L_i}(x)$
replaced by $I_{L_i}(x)$.  For real potentials, 
the $R$-matrix is real, and $\pmb{Z}_O=\pmb{Z}_I^{*}$.  In that case, 
the collision matrix is unitary.  This is not true for complex potentials, 
where the collision matrix remains symmetric, but is not unitary.  Notice 
that the dimension of the $R$ matrix is $N_c$, whether some channels may be 
closed or not.  In contrast, the dimension of matrices $\pmb{Z}_I,\pmb{Z}_O$ 
and $\pmb{U}$ depends on energy, since closed channels are of course absent.  
An equivalent $R$ matrix, limited to open channels, can be defined \cite{LT58,HSV98} 
by explicitly separating open and closed channels in the inversion of 
matrix $\pmb{C}$.  In the present code, we invert the full matrix 
$\pmb{C}$ and select open channels in the definition of matrices $\pmb{Z}_I$ and 
$\pmb{Z}_O$.  Both techniques are of course strictly equivalent.

The theory presented in this subsection is valid for any choice of basis 
functions $\varphi_n(r)$. They can have any normalization, and can be orthogonal or not.
For example, Gaussian functions are intensively used in variational calculations \cite{HKK03,MBH13}. 
The only requirement is that they must be consistently used in (\ref{eq11}) and (\ref{eq12}).
In the following, we use more specifically Lagrange functions.

Notice that, although the $R$-matrix method has been essentially developed for scattering
states, it can be also used for bound states, when the asymptotic behaviour of the wave function plays
an important role \cite{BD83}.

\subsection{Lagrange meshes}
The main advantage of Lagrange functions \cite{Ba15} is that, if the Gauss approximation is 
used for the numerical quadrature, no integral is needed for the matrix 
elements in (\ref{eq11}).  The combination of the $R$-matrix theory with Lagrange 
functions has been already presented and widely used \cite{BHS98,HSV98,PN99,DBD10,DB10}.  
We remind here the main properties of Lagrange functions for their use in the \texttt{rmatrix}
package.

The Lagrange basis functions are always associated with a Gauss quadrature.  
For a finite interval $[0,a]$, the $N$ basis functions, used for all channels, are defined as
\begin{eqnarray}
\varphi_n (r)=(-1)^{N+n}\biggl( \frac{r}{a x_n} \biggr) ^{\nu} \sqrt{ax_n(1-x_n)}\,
\frac{P_N(2r/a-1) }{r-a x_n},
\label{eq15}
\end{eqnarray}
where $P_N(x)$ is a Legendre polynomial of degree $N$, and $x_n$ are the zero's of
\begin{eqnarray}
P_N(2x_n-1)=0.
\label{eq16}
\end{eqnarray}
The $x_n$ values are therefore comprised between 0 and 1.
The factor $(r/ax_n)^{\nu}$, where $\nu$ is a positive integer, is aimed at regularizing the basis
functions at the origin. We use $\nu=1$ unless otherwise mentioned.
The normalization of (\ref{eq15}) is chosen in such a way that the Lagrange condition
\begin{eqnarray}
\varphi_n(ax_m)=\frac{1}{\sqrt{a\lambda_n}}\delta_{nm}
\label{eq17}
\end{eqnarray}
is satisfied.  In this equation, $\lambda_n$ is the weight of the Gauss-Legendre 
quadrature associated with the $[0,1]$ interval.

Functions (\ref{eq15}) can be used as any basis, i.\ e.\ with a standard calculation of
the integrals.  However, the efficiency of the Lagrange 
method is exploited at best when using the associated Gauss approximation for the 
quadratures.  Within this approximation, the overlap and potential matrix elements read
\begin{eqnarray}
&&\langle \varphi_n|\varphi_m\rangle =\int_0^a \varphi_n(r)\varphi_m(r) dr\approx \delta_{nm}, \\
&&\langle \varphi_n|V|\varphi_m\rangle =\int_0^a \varphi_n(r)V(r)\varphi_m(r) dr \approx
V(a x_n)\delta_{nm}. \nonumber
\label{eq18}
\end{eqnarray}
The matrix elements of the kinetic energy are available in an analytical 
form \cite{BGS02}. Within the Gauss approximation, they read
\begin{eqnarray}
\langle \varphi_n \vert T+{\cal L}(B)\vert \varphi_n \rangle \approx \frac{\hbar^2}{2\mu a}
\bigl[ \varphi_n(a)\bigr]^2
\biggl[
\frac{(4N^2+4N+3) x_n(1-x_n) - 6 x_n + 1}{3x_n (1-x_n)}-B \biggr],
\label{eq19}
\end{eqnarray}
and 
\begin{eqnarray}
\langle \varphi_n \vert T+{\cal L}(B)\vert \varphi_m \rangle&\approx&\frac{\hbar^2}{2\mu a}\,
\varphi_n(a) \varphi_m(a) \nonumber \\
&& \times 
\biggl[ N^2+N+1 + \frac{x_n+x_m-2x_nx_m}{(x_n-x_m)^2} -\frac{1}{1-x_n}-\frac{1}{1-x_ m}-B \biggr] {\rm \ for\ }n\ne m.
\label{eq19b}
\end{eqnarray}
In these equations, $\varphi_n(a)$ takes the simple form
\begin{eqnarray}
\varphi_n(a)=(-1)^{N+n}\sqrt{\frac{1}{ax_n(1-x_n)}},
\label{eq25b}
\end{eqnarray}
where (\ref{eq15}) has been used with $\nu=1$.
Consequently, all matrix elements can be computed very easily. The remarkable accuracy
of the Lagrange-mesh method, even with approximate matrix elements, has been discussed in
the literature \cite{BHV02,Sl11}. 

The direct calculation of matrix elements can 
be even extended to non local potentials.  A non-local potential $W(r,r')$ is defined from
\begin{eqnarray}
Wu=\int W(r,r')u(r')dr'.
\label{eq20}
\end{eqnarray}
Non-local potentials 
naturally arise from exchange terms in microscopic methods \cite{Ho77}, but
raise important numerical difficulties in finite-difference methods.
A matrix element between two Lagrange functions is obtained from the simple 
expression
\begin{eqnarray}
\langle \varphi_n|W|\varphi_m\rangle \approx a \sqrt{\lambda_n \lambda_m} W(a x_n,a x_m),
\label{eq21}
\end{eqnarray}
where, as for local potentials, no quadrature is required.  The presence of a non-local potential 
can be easily accounted for in the $R$-matrix method by including matrix elements 
(\ref{eq21}) in the $\pmb{C}$ matrix (\ref{eq11}).

\subsection{Wave functions}
In the internal region, coefficients $f^{\omega}_{in}$ [see eq.\ (\ref{eq5})] are obtained 
from Eq.\ (\ref{eq7}) as
\begin{eqnarray}
f^{\omega}_{in}=\sum_{j,m}\bigl(\pmb{C}^{-1}\bigr)_{in,jm}
\langle \varphi_m \vert {\cal L}_j \vert u^{\omega}_{j,{\rm ext}}
\rangle,
\label{eq22}
\end{eqnarray}
which is valid for open as well as for closed channels.  
These coefficients permit the calculation of the internal wave function at 
any $r$ value, as soon as the collision matrix, and hence the external wave function,
is known.  At the mesh points, the wave functions take the simple value
\begin{eqnarray}
u^{\omega}_{i,{\rm int}}(ax_n)=\frac{1}{\sqrt{a\lambda_n}}f^{\omega}_{in},
\label{eq23}
\end{eqnarray}
where we have used the Lagrange condition (\ref{eq17}).  Coefficients (\ref{eq22}) 
are therefore, up to a scaling factor, the wave function at the mesh points.

Matrix elements of an operator $O$ are determined from
\begin{eqnarray}
\langle u^{\omega}_{i,{\rm int}} \vert O \vert u^{\omega}_{j,{\rm int}} \rangle &=&
\sum_{n,m}f^{\omega}_{in}f^{\omega}_{jm} \langle \varphi_n \vert O \vert\varphi_m \rangle \nonumber \\
&\approx &\sum_n f^{\omega}_{in}f^{\omega}_{jn} O(ax_n). 
\label{eq24}
\end{eqnarray}

In the external region, the collision matrix $\pmb{U}$ provides open-channel 
contributions.  For a closed channel $c$, the amplitude $A_{c}^{\omega}$ (\ref{eq4}) is
obtained 
from the continuity of the wave function as
\begin{eqnarray}
A_{c}^{\omega}=\frac{1}{W_{-\eta_c,L_c+1/2}(2k_c a)}
\sum_n f^{\omega}_{cn}\varphi_n(a).
\label{eq25}
\end{eqnarray}

\subsection{Propagation techniques}
In some cases, the fundamental $R$-matrix condition (\ref{eq3}) may be satisfied at 
very large distances only.  This may occur, for example, in three-body 
scattering \cite{TDE00,DTB06}, in the scattering of exotic nuclei where 
higher-order Coulomb multipoles are important \cite{DD12b}, or in electron-atom 
scattering \cite{Bu11}.  In this situation, the use of a large channel radius 
requires many basis functions to expand the internal function (\ref{eq5}) with an acceptable
accuracy.  This problem has been addressed by using 
modified Coulomb functions, taking account of non-monopole terms \cite{RMP80}.  
Drawbacks of this technique are that $(i)$ the symmetry and the unitarity of the collision matrix 
are partly lost, since the Wronskian of the modified Coulomb functions depends
on the channel (see Ref.\ \cite{LT58}); $(ii)$ a backwards propagation is necessary, starting from
distances large enough so that the Coulomb form (\ref{eq3}) is valid.

An alternative, essentially developed in atomic physics, consists in dividing the $[0,a]$ 
interval in smaller intervals, each of them involving a small number of basis 
functions.  In propagation methods \cite{LW76,BBM82}, the $R$-matrix is propagated 
from one interval to the next interval, and the process is repeated until 
the last interval, where $r=a$.

Let us consider $N_s$ intervals $[a_{\alpha-1},a_{\alpha}]$ with $a_0=0$, and 
$a_{N_s}=a$.  In each interval $(\alpha=1,\ldots ,Ns)$, the wave function is expanded as 
\begin{eqnarray}
u_{i}^{\alpha}(r)=\sum_n f^{\alpha}_{in}\varphi^{\alpha}_n(r),
\label{eq26}
\end{eqnarray}
with the relationships
\begin{eqnarray}
&&u_{i}^{1}(0)=0, \nonumber \\
&&u_{i}^{\alpha}(a_{\alpha})=u_{i}^{\alpha+1}(a_{\alpha}), \nonumber \\
&&u_{i}^{N_S}(a)=u_{i,{\rm ext}}(a),
\label{eq27}
\end{eqnarray}
where indices $\omega$ (entrance channel) and ``int" (internal wave function) are implied.

A Bloch operator ${\cal L}_{\alpha}(B_i)$ is now defined at each boundary as
\begin{eqnarray}
{\cal L}_{\alpha}(B_i)=\frac{\hbar^2}{2\mu}  \delta (r-a_{\alpha}) 
\biggl( \frac{d}{dr} -\frac{B_{i\alpha}}{r}\biggr).
\label{eq28}
\end{eqnarray}
We use the outwards propagation, where the $R$-matrix at $r=a_{\alpha}$ is obtained from the $R$ 
matrix at $r=a_{\alpha-1}$ as
\begin{eqnarray}
a_{\alpha}\pmb{R}(a_{\alpha})=\pmb{\cal R}^{\alpha}_{00}-\pmb{\cal R}^{\alpha}_{01}\bigl[
\pmb{\cal R}^{\alpha}_{11}+a_{\alpha-1}\pmb{R}(a_{\alpha-1})\bigr]^{-1}\pmb{\cal R}^{\alpha}_{10}.
\label{eq29}
\end{eqnarray}
In this equation, matrix $\pmb{\cal R}^{\alpha}_{\beta \gamma}$ is given by
\begin{eqnarray}
\bigl( \pmb{\cal R}^{\alpha}_{\beta \gamma})_{ij}=\frac{\hbar^2}{2\mu}
\sum_{n,m}\varphi^{\alpha}_{n}(a_{\alpha-\beta})
\bigl(\pmb{C}^{\alpha}\bigr)^{-1}_{in,jm}
\varphi^{\alpha}_{m}(a_{\alpha-\gamma}),
\label{eq30}
\end{eqnarray}
where $\beta$ and $\gamma$ take the values $(0,1)$, and are associated with the
limits of subinterval $\alpha$.  Matrix $\pmb{C}^{\alpha}$ is a direct generalization of 
(\ref{eq11}) as
\begin{eqnarray}
C^{\alpha}_{in,jm}=\langle \varphi_n \vert
(T_i +{\cal L}_{\alpha}(B_i)-{\cal L}_{\alpha-1}(B_i)+E_i-E) \delta_{ij}+V_{ij} \vert \varphi_m \rangle_{\alpha},
\label{eq30b}
\end{eqnarray}
where the Dirac notation means that the integration is performed over the $[a_{\alpha-1},a_{\alpha}]$
interval.
The extension of Lagrange functions, and the relevant matrix elements 
are given in \cite{DB10}. In particular, it is stressed that two different regularization $\nu$
values must be used. In the first subinterval $[0,a_1]$, $\nu=1$ to ensure that the wave function vanishes at the
origin, and $\nu=0$ in the other subintervals.

The collision matrix is obtained from Eqs.\ (\ref{eq13}-\ref{eq14}) with the $R$ 
matrix (\ref{eq29}) at the last subinterval, where $a_{N_s}=a$.  The propagation method therefore replaces a $N\times N$ 
problem by $N_s$ smaller $N'\times N'$ problems.  In practice, the inversion of 
matrix $\pmb{C}$ [\ref{eq12}, \ref{eq30}] is the main part of the numerical calculation.  
As the inversion computer time roughly increases with $N^3$, the propagation technique reduces 
this time by about $N_s^2$.

In each subinterval, the wave function (\ref{eq26}) is given by
\begin{eqnarray}
u^{\alpha}_i(r)=\sum_{n,j,m}\bigl(\pmb{C}^{\alpha}\bigr)^{-1}_{in,jm}
\, \langle \varphi^{\alpha}_n \vert
{\cal L}_{\alpha}(B_i)-{\cal L}_{\alpha-1}(B_i)\vert u^{\alpha}_j \rangle
\, \varphi^{\alpha}_m(r).
\end{eqnarray}
Then we have, in vector notation,
\begin{eqnarray}
{\pmb u}^{\alpha}(a_{\alpha-1})=
\pmb{\cal R}^{\alpha}_{10} {\pmb u}^{\alpha \prime}(a_{\alpha})-
\pmb{\cal R}^{\alpha}_{11} {\pmb u}^{\alpha \prime}(a_{\alpha-1})
=a_{\alpha-1} \pmb{R}(a_{\alpha-1})
{\pmb u}^{\alpha \prime}(a_{\alpha-1}).
\end{eqnarray}
This equation provides
\begin{eqnarray}
{\pmb u}^{\alpha \prime}(a_{\alpha-1})=\bigl[ a_{\alpha-1}{\pmb R}(a_{\alpha-1})+\pmb{\cal R}^{\alpha}_{11}
\bigr]^{-1}
\pmb{\cal R}^{\alpha}_{10} {\pmb u}^{\alpha \prime}(a_{\alpha}),
\label{eq31}
\end{eqnarray}
which links the derivatives at the limits of the subinterval $\alpha$.  
Knowing the collision matrix, we start with $\alpha=N_s$ and
\begin{eqnarray}
u^{N_S \prime}_i(a)=u^{\prime}_{i, {\rm ext}}(a),
\label{eq32}
\end{eqnarray}
where $u_{i, {\rm ext}}(r)$ is given by (\ref{eq4}).  Then, $\alpha=N_{s-1},\ldots,1$, and the 
derivatives are determined by a backwards recurrence over all $a_{\alpha}$ values.  
Finally, coefficients $f^{\alpha}_{jm}$ are obtained from the linear system
\begin{eqnarray}
\sum_{j,m}f^{\alpha}_{jm}C^{\alpha}_{in,jm}=\langle 
\varphi^{\alpha}_n \vert {\cal L}_{\alpha}(B_i)-{\cal L}_{\alpha-1}(B_i)\vert
u_{i}^{\alpha} \rangle,
\label{eq33}
\end{eqnarray}
which is solved in each subinterval $\alpha$.  

Notice that the propagation method is not directly applicable to non-local potentials, since they
require integrals starting from zero.

\section{Description of the package}
\label{manual}
The package consists in two parts: three subroutines which must be called by the user, and 
additional general-purpose subroutines. The list of these subroutines is given in 
Sect. \ref{sec_list}. 

As mentioned previously, the inversion of matrix $\pmb{C}$ represents the main part of the computer 
time. Consequently, the inversion subroutines \texttt{cminv\_sym} (symmetric matrices)
and \texttt{cminv\_nsym} (non-symmetric matrices) contains two options. The first option is
to use the subroutine \texttt{cmatinv} included in the package. The second option is to
use subroutines from the LAPACK library \cite{La92}:  \texttt{zsytrf} and  \texttt{zsytrf} (symmetric matrices), or \texttt{zgetrf} and  \texttt{zgetri} (non-symmetric matrices). This second option should be considered for 
many-channel calculations, in particular when a multi-CPU environment is available.
The user can choose the option by commenting out the unnecessary lines in the subroutines
\texttt{cminv\_sym} and \texttt{cminv\_nsym}. 

Units are MeV and fm for energies and lengths, respectively.
All energies are given in the center-of-mass (c.m.) frame.
 
The three main subroutines are:
\begin{description}
	\item [1. rmat\_ini:] returns the abscissas of the mesh points. 
	\item [2. rmatrix:] the main subroutine. Returns the collision matrix, and (optionally) the wave function.
	It must be called for given energy and spin/parity.
	\item [3. wf\_print:] returns the wave function on a uniform mesh (specified by the user).
\end{description}

The subroutine calling sequences are:
\begin{verbatim}
SUBROUTINE RMAT_INI(NR,NS,RMAX,ZRMA)
IMPLICIT REAL*8(A-H,O-Z)
DIMENSION ZRMA(NS*NR)
\end{verbatim}

Input variables: 
\begin{itemize}
	\item \texttt{NR} (integer*4): number of basis functions per interval $N$.
	\item \texttt{NS} (integer*4): number of intervals $N_s$ ($N_s=1$ if propagation is not used).
	\item \texttt{RMAX} (real*8): $R$-matrix channel radius $a$.
\end{itemize}

Output variables:
\begin{itemize}
	\item \texttt{ZRMA(NR*NS)} (array, real*8): abscissas of the Lagrange mesh, where the potentials
	must be calculated (to be used as input in the subroutine \texttt{rmatrix}).
\end{itemize}

\begin{verbatim}  
SUBROUTINE RMATRIX(NCH,LVAL,QK,ETA,RMAX,NR,NS,CPOT,CU,NCP1,NDIM,NOPEN,TWF,CF,
  NWF1,NWF2,NC,NVC,NCP2,CPNL)
IMPLICIT REAL*8(A,B,D-H,O-Z)
IMPLICIT COMPLEX*16(C)
DIMENSION LVAL(NCH),QK(NCH),ETA(NCH),CPOT(NCP1,NDIM,NDIM),CU(NDIM,NDIM),
  CF(NWF1,NWF2,NC),NVC(NC),CPNL(NCP2,NDIM,NDIM)
LOGICAL TWF
\end{verbatim}
Input variables: 
\begin{itemize}
	\item \texttt{NCH} (integer*4): number of channels (total number, including the various angular
	momenta $L_i$).
	\item \texttt{LVAL(NCH)} (array, integer*4): angular momenta $L_i$ of the channels.
	\item \texttt{QK(NCH)} (array, real*8): wave numbers $k_i$ of the channels.\\
	For a closed channel I, \texttt{QK(I)} should be given as a negative number.
	\item \texttt{ETA(NCH)} (array, real*8): Sommerfeld parameters $\eta_i$ of the channels.
	\item \texttt{RMAX} (real*8): $R$-matrix channel radius $a$ (as entered in \texttt{rmax\_ini}).
	\item \texttt{NR} (integer*4): number of basis functions per $[a_{\alpha-1},a_{\alpha}]$
	interval, $N$ (as entered in \texttt{rmax\_ini}).
	\item \texttt{NS} (integer*4): number of intervals $N_s$ (as entered in \texttt{rmax\_ini}).
	\item \texttt{CPOT(NCP1,NDIM,NDIM)} (array, complex*16): local potentials  $V_{ij}(r)$ divided by $\hbar^2/2\mu$.\\
	  \texttt{CPOT(I,J,K)} refers to the mesh point I and to channels J and K (must be symmetric).
	\item \texttt{NCP1} (integer*4): first dimension of  \texttt{CPOT} as declared in the program where  \texttt{RMATRIX} is called ($\texttt{NCP1}\geq \texttt{NS*NR}$.)
	\item \texttt{NDIM} (integer*4): second and third dimensions of \texttt{CPOT} and \texttt{CU} as 
	declared in the program where  \texttt{RMATRIX} is called ($\texttt{NDIM}\geq \texttt{NCH}$.)
	\item \texttt{TWF} (logical): the wave function is computed if  \texttt{TWF=.TRUE.}
	\item \texttt{NVC(NC)} (array, integer*4): defines the set of entrance channels  (not used if  \texttt{TWF=.FALSE.}).
	\item \texttt{NWF1, NWF2} (integer*4): first and second dimensions of  \texttt{CF} as declared in the program where  \texttt{RMATRIX} is called ($\texttt{NWF1}\geq \texttt{NS*NR}, \texttt{NWF2}\geq \texttt{NCH}$.)
	\item \texttt{NC} (integer*4): number of entrance channels (not used if  \texttt{TWF=.FALSE.}).
	\item \texttt{NCP2} (integer*4): if a non-local potential is present: first dimension of
	\texttt{CPNL}; if not: \texttt{NCP2=0}. For this option, $\texttt{NCP2}\geq \texttt{NR**2}$ 
	and \texttt{NS=1}.)
	\item \texttt{CPNL(NCP2,NDIM,NDIM)}(array, complex*16): non-local potentials $W_{ij}(r,r')$
	divided by $\hbar^2/2\mu$. 	Not used	if  \texttt{NCP2=0}.\\
	\texttt{CPNL(I1*(NR-1)+I2,J,K)} refers to the mesh points I1 and I2, and to channels J and K.
\end{itemize}
\vspace{0.5 cm}
Output variables:
\begin{itemize}
	\item \texttt{CU(NDIM,NDIM)} (array, complex*16): collision matrix and (optionally) amplitudes
	of the closed channels.\\
	 \texttt{CU(1:NOPEN,1:NOPEN)} contains the collision matrix elements associated with the open channels.
	\item \texttt{NOPEN} (integer*4): number of open channels.
\end{itemize}
\vspace{0.5 cm}
If \texttt{TWF=.TRUE.}, the program provides an additional output:
\begin{itemize}
	\item \texttt{CF(NWF1,NWF2,NC)} (array, complex*16): contains the wave function for \texttt{NC} entrance channels.
	Indices of the entrance channels are listed in the array \texttt{NVC}. \\
	 \texttt{CF(I,J,K)} 
	contains the wave function $u_{i,\rm int}^{\omega}(r)$ at the mesh points, for the output channel \texttt{J} and for the entrance channel 
	\texttt{K}. 
	Index \texttt{I} therefore runs from \texttt{I=1} to \texttt{NS*NR}.
	The array \texttt{CF} can be, either used as in Eq.\ (\ref{eq24}), or interpolated to get the wave function at specific $r$ values.
	\item \texttt{CU(NOPEN+1:NCH,1:NOPEN)} contains the amplitudes (\ref{eq25}) of closed channels in the external region
	(complex). Coefficients $f^{\omega}_{in}$ can be determined from Eq. (\ref{eq23}).
\end{itemize}

Let us briefly comment on variables \texttt{NC} and \texttt{NVC(1:NC)}. A ``physical" channel is defined by the
quantum numbers associated with the colliding nuclei (energy, spin, parity, etc.). This means that, if the spins
are different from zero, several $L_{\omega}$ values are associated with that entrance channel, and that the
total wave function involves several $L_{\omega}$ values. For example, let us consider the p+n system, with
$J=1^+$. Although the problem involves a single physical channel, $L_{\omega}=0$ and $L_{\omega}=2$
are necessary to define the total wave function.  

\begin{verbatim}  
SUBROUTINE WF_PRINT(NCH,LVAL,QK,ETA,RMAX,NR,NS,CU,NDIM,NOPEN,CF,
   NWF1,NWF2,ZRMA,IV,NOM,NPOIN,H,CWFTAB)
IMPLICIT REAL*8(A,B,D-H,O-Z)
IMPLICIT COMPLEX*16(C)
DIMENSION LVAL(NCH),QK(NCH),ETA(NCH),CU(NDIM,NDIM),CF(NWF1,NWF2,NOM),
   ZRMA(NS*NR),CWFTAB(NPOIN)
\end{verbatim}
Input variables specific to this subroutine: 
\begin{itemize}
	\item \texttt{IV} (integer*4): exit channel.
	\item \texttt{NOM} (integer*4): entrance channel.
	\item \texttt{NPOIN} (integer*4): number of points where the wave functions is computed.
	\item \texttt{H} (real*8): mesh size.
\end{itemize}	
The output is:
\begin{itemize}
\item \texttt{CWFTAB(NPOIN)} (complex*16): wave function computed at the mesh points (from
the interpolation of vector \texttt{CF})
\end{itemize}

\section{List of additional subroutines}
\label{sec_list}
The additional subroutines are:
\begin{description}
	\item [whit:] computes the Whittaker function  (see Ref. \cite{AS72}).
	\item [legzo:] returns the abscissas and weights of the Gauss-Legendre integration \cite{ZJ96}.
	\item [coulfg:] computes the Coulomb functions \cite{Ba84b}.
	\item [cminv\_sym:] computes the inverse of a symmetric complex matrix; calls either 
	\texttt{cmatinv} (included in the package) or \texttt{zgetrf} and  \texttt{zgetri} (available in the LAPACK library). By default, \texttt{cmatinv} is used. The user can modify this option by
	commenting out a few lines in \texttt{cminv\_sym}.
	\item [cminv\_nsym:] computes the inverse of a non-symmetric complex matrix; calls either 
	\texttt{cmatinv} (included in the package) or \texttt{zsytrf} and  \texttt{zsytri} (available in the LAPACK library). The choice is done as before.
	\item [cmatinv:] computes the inverse of a general complex matrix.
	\item [spline, splint:] used for the interpolation of the wave function \cite{PFT86}.
		
\end{description}

\section{Examples}
\label{sec_exam}
\subsection{General presentation}
We provide examples to run the package in various conditions. 
Each example (here example 1) includes

\begin{itemize}
	\item the main program (calling subroutine  \texttt{RMATRIX})  \texttt{example1.f}.
	\item the $R$-matrix package  \texttt{rmatrix.f}.
	\item the input file  \texttt{data1}.
	\item the output file  \texttt{output1.txt} and the wave function file  \texttt{wf1.txt} are also provided as a test.
\end{itemize}

Installation (we assume the INTEL Fortran compiler \texttt{ifort}):
\begin{enumerate}
\item Compile \texttt{rmatrix.f} \\
\texttt{>ifort -c rmatrix.f}
\item Compile the example and link to \texttt{rmatrix.o} \\
\texttt{>ifort example1.f -o example1.out rmatrix.o}\\
Notice that this line may contain an additional link to the LAPACK library (see Ref.\ \cite{La92} for
the source files and the installation of the library).
\item Run the example \\
\texttt{>./example1.out < data1}
\item The output can be compared with the content of the file  \texttt{output1.txt}. Of course, the main program and the input file can be adapted by the user.
\end{enumerate}	

In each case, the main program (here  \texttt{example1.f}) contains
the potential, and reads some input:
\begin{enumerate}
	\item \texttt{L} (or \texttt{J}), \texttt{NR, NS, RMAX}: as defined before (exit if \texttt{L}$<$0)
	\item \texttt{NE, E0, ESTEP}  \ \\
	\texttt{NE}=number of energies (the line is repeated until \texttt{NE=0}, and goes back to line 1)\\
	\texttt{E0}=first energy \\
	\texttt{ESTEP}=energy step.
\end{enumerate}
If  \texttt{RMAX$<$0}:  \texttt{TWF=.TRUE.}, and the program prints the wave function in a file called  
\texttt{wave\_function.txt} (the absolute value of \texttt{RMAX} is used in the $R$-matrix calculation).

\subsection{Nucleus-nucleus potentials}
For the nuclear interaction, we usually use a Wood-Saxon potential depending on the nucleus-nucleus coordinate
$r$ as
\beq
V_N(r)=-V_0 f(r,R_R,a_R)-i W_0f(r,R_I,a_I),
\label{eq_vn}
\eeq
with
\beq
f(r,R,a)=\frac{1}{1+\exp((r-R)/a)},
\label{eq_ws}
\eeq
and where the amplitudes $(V_0,W_0)$, the ranges $(R_R,R_I)$ and the diffusenesses $(a_R,a_I)$
are parameters, in general fitted to some experimental data. The values adopted in the examples are
taken from the literature, or are chosen as ``reasonable" values to illustrate the use
of the program. They are not intended to fit any experimental data.

For the Coulomb interaction, we use, either the point-like approximation
\beq
V_C(r)=\frac{Z_1Z_2e^2}{r},
\label{eq_coul1}
\eeq
or the more general definition, associated with an uniformly charged sphere
\beq
V_C(r)&=&\frac{Z_1Z_2e^2}{2R_C}\bigl(3-(r/R_C)^2\bigr)  {\rm \ for \ } r\le R_C,\nonumber \\
&=&\frac{Z_1Z_2e^2}{r} {\rm \ for \ } r\ge R_C,
\label{eq_coul2}
\eeq
where $R_C$ is the sphere radius.

\subsection{Numerical conditions}
In our examples, we use
\beq
e^2&=&1.44 {\rm \ MeV.fm},\nonumber \\
\frac{\hbar^2}{2\mu}&=&\frac{\hbar^2}{2m_N \mu_0},\nonumber \\
\frac{\hbar^2}{2m_N}&=&20.736 {\rm \ MeV} \cdot {\rm fm}^2,
\eeq
where the dimensionless reduced mass $\mu_0$ is defined from the (integer) nucleon numbers. This can be changed by the user.

Let us briefly discuss the choice of the $R$-matrix parameters (see also Ref.\ \cite{DB10}):
\begin{itemize}
	\item The channel radius \texttt{RMAX} must be chosen such as the total potential reduces to the monopole
	Coulomb term [see Eq.\ (\ref{eq3})]. For light systems, typical values are $\approx 8-12$ fm. For heavier systems,
	the range of the nuclear potential is larger, and \texttt{RMAX} takes typically values $\approx 10-15$ fm.
	These values are directly associated with the range of the nuclear interaction.
	The channel radius can be, however, significantly larger (typically up to $\approx 100-200$ fm)
	in CDCC calculations involving dipole Coulomb couplings (see Ref. \cite{DD12b} for example).
	\item When \texttt{RMAX} is defined, the number of basis functions \texttt{NS*NR} must be large enough to accurately 
	describe the internal wave functions, and to ensure the matching with the external wave function
	at $r=a$ [see Eq.\ (\ref{eq9})]. Of course this number depends on \texttt{RMAX}: increasing \texttt{RMAX} requires a simultaneous
	increase of the number of basis functions. Typically 5 basis functions per fm is a reasonable
	initial guess.
	\item The use of the propagation method \texttt{(NS}$>$1) allows to split the internal region in
	 \texttt{NS} pieces. This means that, instead of inverting a matrix of size \texttt{(NS*NR*NCH)}$\times$\texttt{(NS*NR*NCH)}, we must invert
	NS matrices of size \texttt{(NR*NCH)}$\times$\texttt{(NR*NCH)}. As the inversion time roughly increases with 
	$N^3$ (where $N$ is the size of the matrix), a substantial benefit can be obtained. The limitation,
	however, is that in each interval, the number \texttt{NR} should be large enough to describe the physical
	wave function. In practice, the lower limit is of the order of \texttt{NR}$\approx$5. A more detailed
	discussion will be presented after the examples.
\end{itemize}	

Of course the collision matrix and the associated wave function should not depend on these parameters, provided they are properly adapted to the physics of the problem. Another severe test is to control the
logarithmic derivative at the channel radius (see Ref.\ \cite{DB10}). In practice, the choice
of the $R$-matrix parameters is guided by a compromise between the accuracy of the calculation,
and computer times as low as possible. These different issues are discussed and illustrated in
Ref.\ \cite{DB10}, and we refer the reader to that reference for more detail.
In the examples given below, the computer times are very short, and do not represent any limitation.
In calculations involving many channels, however, optimizing the choice of the $R$-matrix
parameters is crucial.

\subsection{Example 1: single-channel optical potential \apb}
We illustrate single-channel optical potentials with the \apb\ potential of Goldring {\sl et al.}
\cite{GSW70}. This optical potential is given by Eq. (\ref{eq_vn})
with
\beq
&&V_0=100{\rm \ MeV}, W_0=10{\rm \ MeV}, \nonumber \\
&&R_R=R_I=1.1132\times (208^{1/3}+4^{1/3}) {\rm \ fm},\nonumber \\
&&a_R=a_I=0.5803{\rm \ fm}.
\eeq
For the Coulomb potential, the point-like definition (\ref{eq_coul1}) is used.

The input data file \texttt{data1} contains
\begin{verbatim}
20 60 1 14.00
5 10.0 10.0
0 0 0
20 15 5 14.00
5 10.0 10.0
0 0 0
20 15 5 -16.00
5 10.0 10.0
0 0 0
-1 0 0 0
\end{verbatim}
Three calculations are performed for $L=20$, and for 5 energies (from 10 to 50 MeV by step of 10 MeV). 
The calculations differ by the $R$-matrix parameters. Lines 1 and 2 of the input file correspond to 
\texttt{NS=1, NR=60, RMAX=14} fm
(propagation is not used).
With lines 4 and 5, we illustrate the sensitivity with respect to the mesh points (here \texttt{NS}=5 and \texttt{NR}=15),
and the same  \texttt{RMAX} is used. In the last part (lines 7 and 8), we change \texttt{RMAX}=16 fm, and the program prints the wave functions. 
As mentioned before, these calculations are very fast, and the use of propagation is presented as an
application of the method. 

The output file and a figure with the wave functions are given in Appendix A.

\subsection{Example 2: two-channel nucleon-nucleon potential}
This example deals with a two-channel calculation, with the (real) nucleon-nucleon potential
of Reid \cite{Re68}. We choose the $T=0$ (neutron-proton) soft-core variant (Eqs. (28)-(30) of
Ref.\ \cite{Re68}), which involves central, spin-orbit, and tensor terms.

The input data file \texttt{data2} contains
\begin{verbatim}
1 60 1 7.0
4 12.0 12.0
0 0 0
1 30 2 7.0
4 12.0 12.0
0 0 0
1 25 3 -8.0
4 12.0 12.0
0 0 0
-1 0 0 0
\end{verbatim}
and corresponds to the $J=1^+$ partial wave, involving a mixing of $S$ and $D$ waves.
Again we illustrate the sensitivity against variations of the $R$-matrix parameters
($a=7$ and 8 fm, and different \texttt{(NS,NR)} values).
The collision matrix is parametrized as
\begin{equation}
U^{1^+}_{ij}(E)=
\begin{pmatrix}
\eta_{11}\exp(2i\delta_{11})	& \eta_{12}\exp(2i\delta_{12}) \\ 
\eta_{21}\exp(2i\delta_{21})	& \eta_{22}\exp(2i\delta_{22})
\end{pmatrix}, 
\end{equation}
where channels 1 and 2 correspond to the $S$ and $D$ waves, respectively. Owing to the symmetry and unitary properties for a real potential, a two-channel problem is characterized by 3
independent parameters. 
The printed values are $\delta_{11}$, $\delta_{22}$ and $\eta_{12}$ (corresponding to $\rho_1$ in the notations
of Ref. \cite{Re68}). They can be compared with phase shifts presented in Table VI of Ref. \cite{Re68}.

The output file and a figure with the wave functions are given in Appendix B.

\subsection{Example 3: two-channel optical potential \oca }
In this example, we consider the coupled-channel problem \oca$(0^+,2^+)$ which was
studied in Refs. \cite{RMP80,RMP80b}. When the target remains in the $0^+$ ground state,
a physical channel is defined by the spin of the projectile $I$, and by the relative angular moment 
$\vert J-I \vert \leq L \leq J+I$ (only even or odd values are present, according to the total parity).
In the framework of the rotational model, states belonging to a rotational band can be
characterized by deformation parameters $\beta_{\lambda}$. Considering a single $\lambda$ value,
Ref.\ \cite{RMP80} defines the coupling potential between channel $(IL)$ and $(I'L')$ as 
\beq
V^{J\pi}_{IL,I'L'}(r)=V_N(r)\delta_{II'}\delta_{LL'}-i^{L'-L}A_{\lambda}(I,I',L,L',J)\beta_{\lambda}R_v
\frac{dV_N}{dr},
\eeq
with
\begin{equation}
A_{\lambda}(I,I',L,L',J)=(-1)^{J+\lambda}\frac{\hat{I}\hat{I}'\hat{L}\hat{L}'\hat{\lambda}}
{(4\pi)^{1/2}}
\begin{pmatrix}
I & I' & \lambda  \\ 
0 &  0 & 0
\end{pmatrix} 
\begin{pmatrix}
L & L' & \lambda  \\ 
0 &  0 & 0
\end{pmatrix} 
\begin{Bmatrix}
I & L & J  \\ 
L' &  I' & \lambda
\end{Bmatrix} .
\end{equation}
The symbol $\hat{x}$ stands for $\hat{x}=(2x+1)^{1/2}$, and $R_v$ is the radius of the excited nucleus.

We adopt the parameters of Ref. \cite{RMP80b}, i.e. we consider $J=30^+$, and
use the nuclear potential (\ref{eq_vn}) with
\beq
&&V_0=110{\rm \ MeV}, W_0=20{\rm \ MeV}, \nonumber \\
&&R_R=R_I=1.2\times (44^{1/3}+16^{1/3}) {\rm \ fm},\nonumber \\
&&a_R=a_I=0.5{\rm \ fm}.
\eeq
For the Coulomb interaction, we 
use (\ref{eq_coul2}) with $R_C=R_R$.

We include the $^{44}$Ca ground state ($I=0^+$) and the first excited state 
($I=2^+,E=1.156$ MeV). 
The deformation parameter is $\beta_2=0.4$, with $R_v=1.2\times 44^{1/3}$.
This example was considered in Ref. \cite{RMP80b} to illustrate various iterative methods,
based on the Numerov algorithm. It was emphasized that, although these methods allow to speed up
the calculations, essentially by focusing on a specific element of the collision matrix, they may
raise numerical instabilities when the coupling potentials increase. One of the advantages of the
$R$-matrix method is that, except for possible adaptations of the channel radius, its application
does not depend on the amplitude of the coupling potentials.

The input data file \texttt{data3} contains
\begin{verbatim}
30 25 4 12.0
0 0
2 34.0 10.0
0 0 0
30 25 4 13.0
0 0
2 34.0 10.0
0 0 0
30 25 4 14.0
0 0
2 34.0 10.0
0 0 0
30 50 2 -14.0
10 2.0
2 34.0 10.0
0 0 0
-1 0 0 0
\end{verbatim}

We illustrate calculations with different channel radii and different meshes.
The output file and a figure with the wave functions are given in Appendix C.
We also determine the wave function on a uniform mesh (from 2 fm to 20 fm by step of 2 fm)
by using the subroutine \texttt{wf\_print}.
At $\ecm=44$ MeV, the collision matrix amplitude for ($I'=2,L'=30$) is 
$\vert U^{30^+}_{0,30;2,30}\vert =0.2085$,
in agreement with Ref.\ \cite{RMP80b} (0.2072). Slight differences may occur from different choices of the
physical constants.

\subsection{Example 4: multi-channel optical potential \ac\ with closed channels}
In this example, we aim at illustrating a reaction involving closed channels. We consider
the \ac $(0^+,2^+,4^+)$ system, with the potential defined in example 3 and the parameters
\beq
&&V_0=110{\rm \ MeV}, W_0=20{\rm \ MeV}, \nonumber \\
&&R_R=R_I=1.2\times (4^{1/3}+12^{1/3}) {\rm \ fm},\nonumber \\
&&a_R=a_I=0.5{\rm \ fm}.
\eeq
We include
the $2^+$ (4.44 MeV) and $4^+$ (14.40 MeV) states with $\beta_2=0.58$ \cite{RNT01}.
Notice that our goal is just to provide a numerical example, and the conditions of the
calculation are not intended to fit any data.

For $J=3^-$ we have $L=3$ for $I=0^+$,  $L=1,3,5$ for $I=2^+$ and  $L=1,3,5,7$ for $I=4^+$.
The calculation therefore involves 8 channels, but the size of the collision matrix of course depends
on energy. In Appendix D, we show 3 components of the wave function for $E=4$ MeV, where the \ac $(2^+)$
and \ac $(4^+)$ channels are closed.

The input data file \texttt{data4} contains
\begin{verbatim}
3 25 4 09.0
5 4.0 4.00
0 0 0
3 25 4 10.0
5 4.0 4.00
0 0 0
3 20 4 -11.0
5 4.0 4.00
0 0 0
-1 0 0 0 
\end{verbatim}

\subsection{Example 5: nucleon-nucleon scattering with a non-local potential}
Here we present a calculation with a non-local potential.
We use the Yamaguchi potential \cite{Ya65}, which was considered in a previous
$R$-matrix calculation \cite{HRB02}. For $L=0$, this non-local potential is defined as
\begin{equation}
W(r,r')=-\frac{\hbar^2}{2\mu}2\beta(\alpha+\beta)^2\exp(-\beta (r+r')),
\label{eq_yam}
\end{equation}
and can be solved analytically \cite{HRB02}. As in Ref. \cite{HRB02}, we take
$\alpha=0.2316053$ fm$^{-1}$, $\beta=1.3918324$ fm$^{-1}$, and $\hbar^2/2\mu=41.472$
MeV fm$^2$.

The input data file \texttt{data5} contains
\begin{verbatim}
10 8.00
2 0.1 9.9
0 0 0
15 8.00
2 0.1 9.9
0 0 0
15 -12.00
2 0.1 9.9
0 0 0
0 0
\end{verbatim}

The phase shifts reproduce the values given in  Ref. \cite{HRB02}. The output and the wave functions
are given in Appendix E.

\section{Discussion of the propagation method}
In this section, we illustrate the propagation method described in Sect. 2.4 with the 
two-channel example 3.
We use an angular momentum $J=30^+$, a channel radius $a=15$ fm, and we compare various conditions
with a reference calculation with $N_R=100$ and $N_S=1$ (without propagation). Then, we split
the interval $[0,a]$ in different ways ($N_S=2$ to 50), and keep the same total number of basis
functions. In other words, we use $N_R=100/N_S$ functions in each subinterval. We consider 3 energies: 
34, 44 and 54 MeV.

\begin{figure}[h]
	\centering
	\includegraphics*[width=.6\textwidth,clip=]{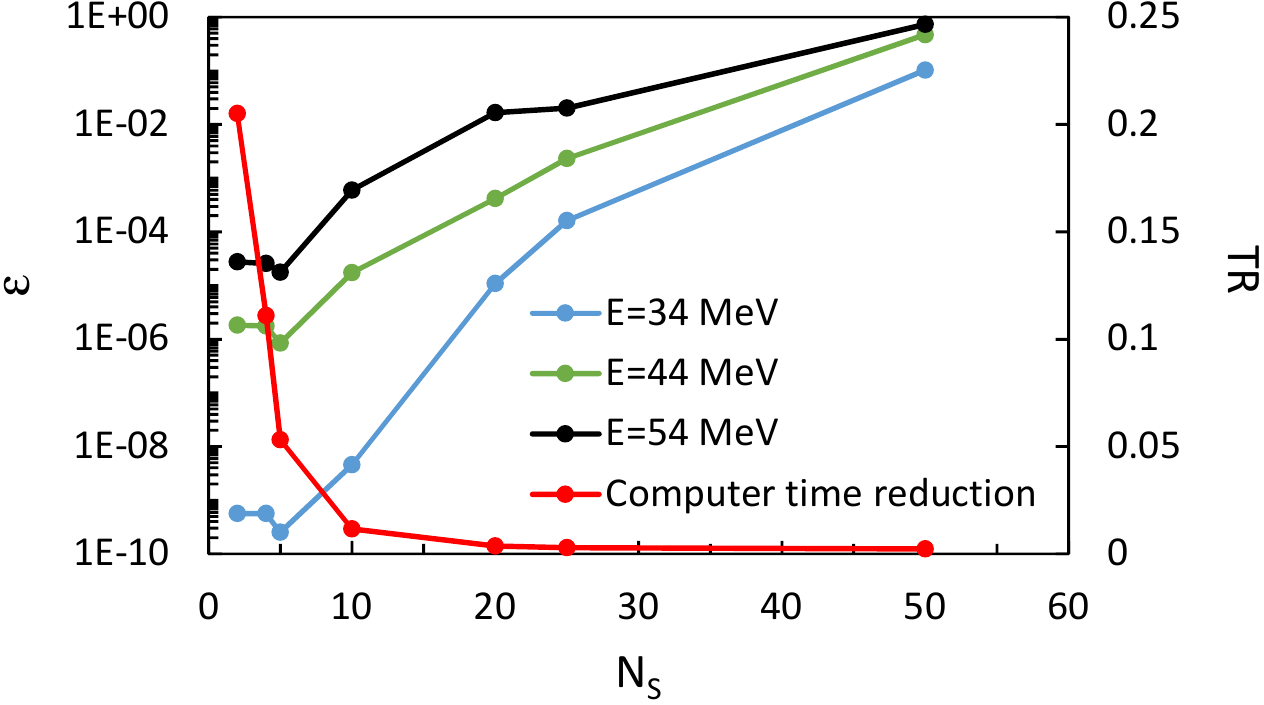}
	\caption{Accuracy (\ref{eq_eps}) (left axis) and computer time reduction (\ref{eq_time}) (right
		axis) with the propagation method. The conditions correspond to
		example 3 ($J=30^+$, $a=15$ fm), and the reference calculation is performed with $N_S=1, N_R=100$ (see text for detail).}
	\label{fig_ns}
\end{figure}

In Fig.\ \ref{fig_ns}, we present the accuracy of the method, and the time reduction for different
values of $N_S$. Of course, increasing $N_S$ reduces the computer time by a factor $\approx N_S^2$, 
as discussed in Sect.\ 2.4. This is confirmed by Fig.\ \ref{fig_ns} (right vertical axis) where we show
the time reduction 
\begin{eqnarray}
TR=\frac{{\rm time}(N_S)}{{\rm time}(1)}.
\label{eq_time}
\end{eqnarray}
However, increasing $N_S$ means that $N_R$ should be decreased if we want to keep a fixed number (100) of
basis functions. Then the accuracy, measured by
\begin{eqnarray}
\epsilon=\biggl| \frac{U_{ij}(N_S)-U_{ij}(1)}{U_{ij}(1)}\biggr|
\label{eq_eps}
\end{eqnarray}
takes unphysical values when $N_R$ is too small or, in other words, when $N_S$ is too large. A fair
compromise is that $N_R\approx 5$ represents a lower limit. Below this value, the number of Lagrange functions
is not large enough to describe accurately the wave function, even in a small subinterval.
We consider here the $ij$ channels where $\epsilon$ takes the largest values, but our conclusions 
are very similar for all elements of the collision matrix.

Of course, the values presented in Fig.\ \ref{fig_ns} should be considered as indicative only. As the conditions may vary with the energy, the angular momentum, the size of the system, the range of the potential,
similar analyses should be performed. However the existence of a compromise, as well as a
reasonable lower limit $N_R\approx 5$, are common to all
conditions of calculations.

\section{Conclusion}
\label{sec_conc}
We propose here an $R$-matrix Fortran package which provides, for a given energy and spin/parity, the collision
matrix and the associated wave function. Then the user may easily compute any cross section by
using well known formula \cite{LT58,DB10,CH13}. Combined with Lagrange functions, the $R$-matrix theory
is an efficient tool to determine scattering properties from the coupling potentials. Compared to
finite-difference methods, it offers two obvious advantages: $(i)$ non-local potentials are
treated on an equal footing, and do not require any specific adaptation; with Lagrange
functions, the matrix elements of a local potential are diagonal, whereas they are non-diagonal for non-local potentials. $(ii)$ Closed channels are included consistently with open channels. 
Possible numerical instabilities, due to the inclusion of closed channels, are absent in the $R$-matrix method.

Single-channel calculations are extremely fast with modern computers.
For large-scale calculations (i.e. from several tens of channels), the inversion of the complex
matrix $\pmb{C}$ (see eqs.\ (\ref{eq12})and (\ref{eq30})) is the main numerical issue. We showed that
the computer time can be significantly reduced by using the propagation method. For the user, the
only change is that the mesh points, where the coupling potentials must be calculated, are different
with and without propagation. This technique is therefore simple to implement in any code. In addition,
we mentioned that using \texttt{LAPACK} subroutines for the matrix inversion may also speed up the calculation, in particular in a multi-CPU environment using \texttt{OPENMP}.

Solving a coupled-channel system in the continuum is a frequent problem in physics. We believe that this Fortran package is simple to use, as soon as the coupling potentials are determined. These potentials may be deduced
from many different models, such as the microscopic RGM method, the CDCC method, standard coupled-channel
theories, or many others. Consequently, the present package may be helpful to many potential users who need a minor adaptation
of their bound-state codes to extend them to continuum states.

\section*{Acknowledgments}
I am very grateful to L.F. Canto, K. Hagino, A. Moro, M. Rodrigues Gallardo and I.J. Thompson for helpful
comments on the manuscript and for their tests of the code.
This text presents research results of the IAP programme P7/12 initiated by the Belgian-state 
Federal Services for Scientific, Technical and Cultural Affairs. 

\section*{Appendix A}
This appendix gives the output of example 1.
\begin{verbatim}
total angular momentum= 20
number of basis functions per interval= 60
number of intervals=  1
channel radius=  14.0000
Number of energies:  5
Initial energy:   10.0000
Energy step:   10.0000
E (MeV)=  10.000   Collision matrix=  1.0000E+00  5.9801E-19
E (MeV)=  20.000   Collision matrix=  1.0000E+00  7.4950E-07
E (MeV)=  30.000   Collision matrix=  9.9893E-01  9.0496E-03
E (MeV)=  40.000   Collision matrix=  6.5081E-01  2.9560E-01
E (MeV)=  50.000   Collision matrix=  6.4367E-02  4.1130E-02
Number of energies:  0
Initial energy:    0.0000
Energy step:    0.0000

total angular momentum= 20
number of basis functions per interval= 15
number of intervals=  5
channel radius=  14.0000
Number of energies:  5
Initial energy:   10.0000
Energy step:   10.0000
E (MeV)=  10.000   Collision matrix=  1.0000E+00  5.9631E-19
E (MeV)=  20.000   Collision matrix=  1.0000E+00  7.4950E-07
E (MeV)=  30.000   Collision matrix=  9.9893E-01  9.0496E-03
E (MeV)=  40.000   Collision matrix=  6.5081E-01  2.9560E-01
E (MeV)=  50.000   Collision matrix=  6.4367E-02  4.1130E-02
Number of energies:  0
Initial energy:    0.0000
Energy step:    0.0000

total angular momentum= 20
number of basis functions per interval= 15
number of intervals=  5
channel radius=  16.0000
Number of energies:  5
Initial energy:   10.0000
Energy step:   10.0000
E (MeV)=  10.000   Collision matrix=  1.0000E+00  2.8398E-17
E (MeV)=  20.000   Collision matrix=  1.0000E+00  4.4229E-06
E (MeV)=  30.000   Collision matrix=  9.9879E-01  1.0301E-02
E (MeV)=  40.000   Collision matrix=  6.5059E-01  2.9599E-01
E (MeV)=  50.000   Collision matrix=  6.4381E-02  4.1206E-02
Number of energies:  0
Initial energy:    0.0000
Energy step:    0.0000
\end{verbatim}

\begin{figure}[h]
	\centering
	\includegraphics*[width=.6\textwidth,clip=]{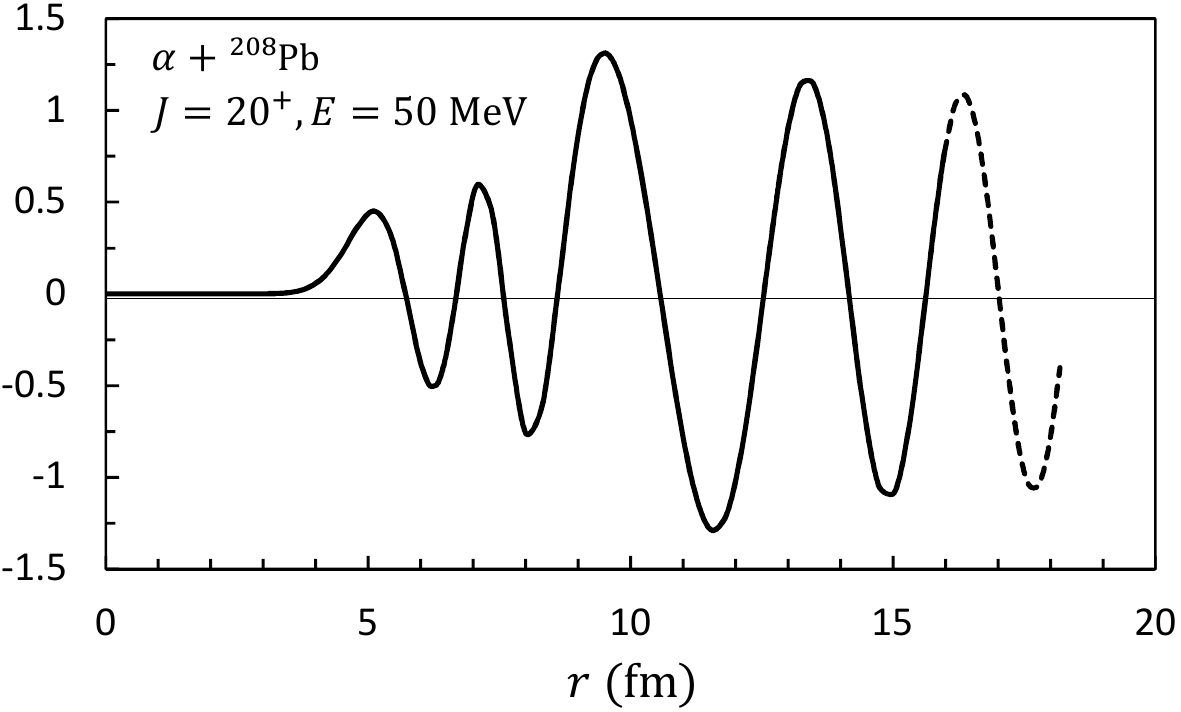}
	\caption{Real part of the \apb\ $J=20^+$ wave function at $\ecm=50$ MeV, with a channel radius $a=16$ fm.
		The solid and dashed lines represent the internal and external wave functions, respectively.}
	\label{fig1}
\end{figure}

In Fig. \ref{fig1}, we show the real part of the wave function at 50 MeV. The internal part
(solid line) is computed with Eq.\ (\ref{eq5}), and the external part (dashed line) with 
Eq.\ (\ref{eq4b}). The quality of the matching at the channel radius ($a=16$ fm) supports
the accuracy of the calculation. Another test is provided by the stability of the collision matrix against
variations of the channel radius and of the basis size.

\section*{Appendix B}
\begin{figure}[h]
	\centering
	\includegraphics*[width=.6\textwidth,clip=]{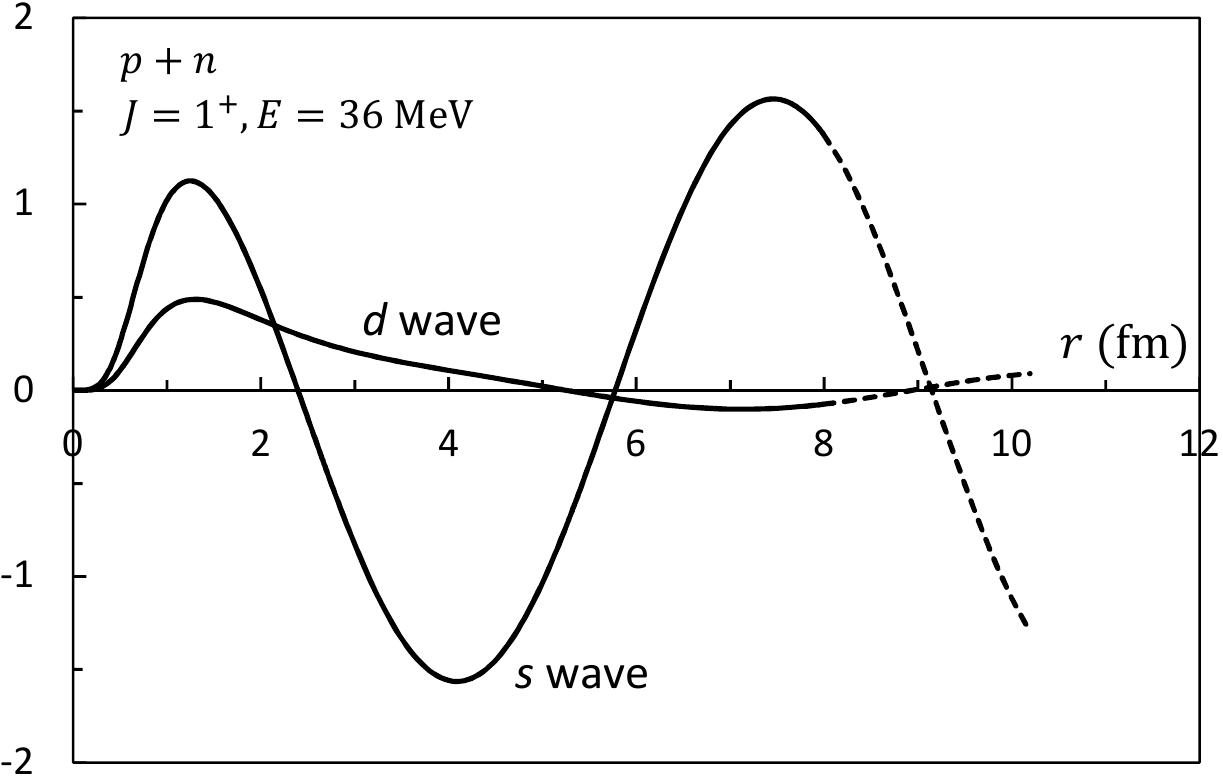}
	\caption{Real part of the $J=1^+$ $n-p$ wave function at $\ecm=36$ MeV, with a channel radius $a=8$ fm
		(the entrance channel is $L_{\omega}=0$).
		The solid and dashed lines represent the internal and external wave functions, respectively.}
	\label{fig2}
\end{figure}

This appendix gives the output of example 2. The wave function at $\ecm=36$ MeV is shown in Fig.\ \ref{fig2}.
\begin{verbatim} 

total angular momentum=  1
number of basis functions per interval= 60
number of intervals=  1
channel radius=   7.0000
Number of energies:  4
Initial energy:   12.0000
Energy step:   12.0000
E (MeV)= 12.0 phase shift (rad.)=  1.4256E+00 -4.8047E-02 eta_12=  6.7922E-02
E (MeV)= 24.0 phase shift (rad.)=  1.1052E+00 -1.1502E-01 eta_12=  8.2249E-02
E (MeV)= 36.0 phase shift (rad.)=  9.0165E-01 -1.6959E-01 eta_12=  9.9708E-02
E (MeV)= 48.0 phase shift (rad.)=  7.4889E-01 -2.1425E-01 eta_12=  1.1575E-01
Number of energies:  0
Initial energy:    0.0000
Energy step:    0.0000

total angular momentum=  1
number of basis functions per interval= 30
number of intervals=  2
channel radius=   7.0000
Number of energies:  4
Initial energy:   12.0000
Energy step:   12.0000
E (MeV)= 12.0 phase shift (rad.)=  1.4256E+00 -4.8047E-02 eta_12=  6.7922E-02
E (MeV)= 24.0 phase shift (rad.)=  1.1052E+00 -1.1502E-01 eta_12=  8.2249E-02
E (MeV)= 36.0 phase shift (rad.)=  9.0165E-01 -1.6959E-01 eta_12=  9.9708E-02
E (MeV)= 48.0 phase shift (rad.)=  7.4889E-01 -2.1425E-01 eta_12=  1.1575E-01
Number of energies:  0
Initial energy:    0.0000
Energy step:    0.0000

total angular momentum=  1
number of basis functions per interval= 25
number of intervals=  3
channel radius=   8.0000
Number of energies:  4
Initial energy:   12.0000
Energy step:   12.0000
E (MeV)= 12.0 phase shift (rad.)=  1.4258E+00 -4.9358E-02 eta_12=  6.4110E-02
E (MeV)= 24.0 phase shift (rad.)=  1.1052E+00 -1.1506E-01 eta_12=  8.2470E-02
E (MeV)= 36.0 phase shift (rad.)=  9.0182E-01 -1.7002E-01 eta_12=  9.7677E-02
E (MeV)= 48.0 phase shift (rad.)=  7.4900E-01 -2.1469E-01 eta_12=  1.1440E-01
Number of energies:  0
Initial energy:    0.0000
Energy step:    0.0000
\end{verbatim}

\section*{Appendix C}
\begin{figure}[h]
	\centering
	\includegraphics*[width=.6\textwidth,clip=]{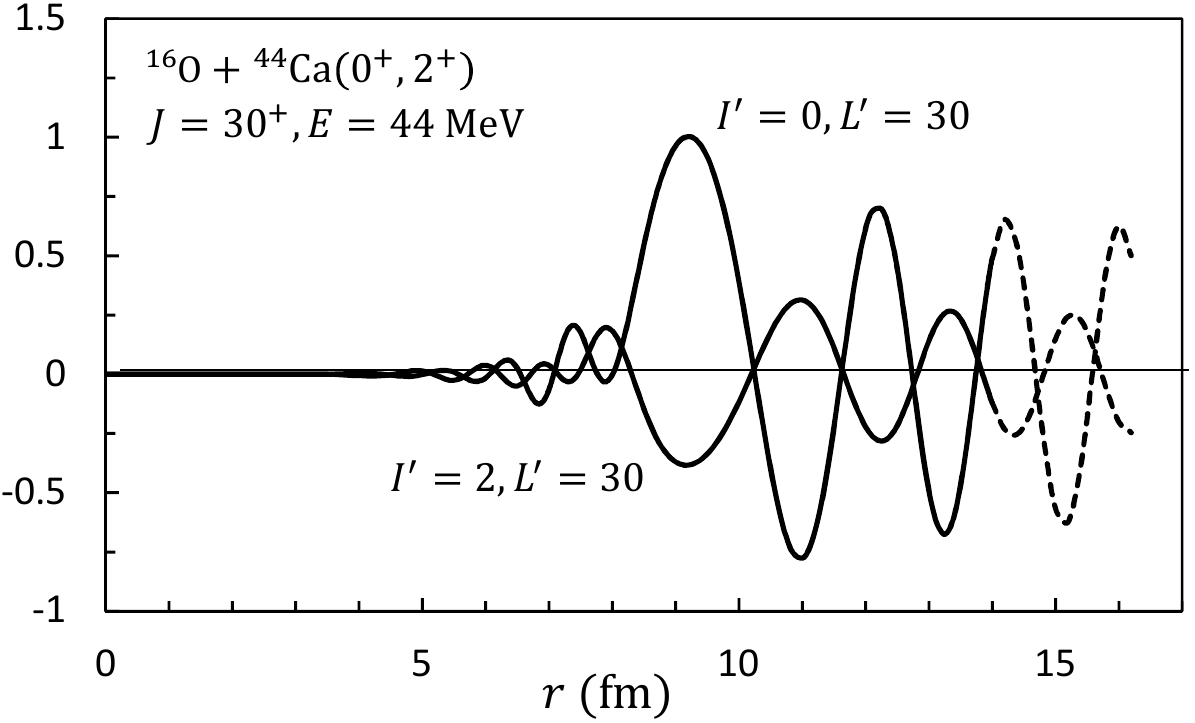}
	\caption{Real part of the $J=30^+$  \oca\ wave function at $\ecm=44$ MeV, with a channel 
		radius $a=14$ fm. We show the elastic ($I'=0,L'=30$) and inelastic  ($I'=2,L'=30$) components.
		The solid and dashed lines represent the internal and external wave functions, respectively.}
	\label{fig3}
\end{figure}

This appendix gives the output of example 3. The wave function at $\ecm=44$ MeV is presented in Fig.\
\ref{fig3}.

\begin{verbatim}   
total angular momentum= 30
number of basis functions per interval= 25
number of intervals=  4
channel radius=  12.0000
Number of energies:  2
Initial energy:   34.0000
Energy step:   10.0000
E (MeV)=  34.000 Amplitude=           9.9408E-01  1.9823E-02  9.9839E-03  6.8090E-03
E (MeV)=  34.000 phase shift (rad.)=  1.3878E-02 -6.5505E-01 -6.6877E-01 -6.7803E-01
E (MeV)=  44.000 Amplitude=           5.3784E-01  1.6169E-01  2.0842E-01  2.1203E-01
E (MeV)=  44.000 phase shift (rad.)=  1.8330E-01  2.4799E-01  3.4390E-02 -7.6706E-02

total angular momentum= 30
number of basis functions per interval= 25
number of intervals=  4
channel radius=  13.0000
Number of energies:  2
Initial energy:   34.0000
Energy step:   10.0000
E (MeV)=  34.000 Amplitude=           9.9373E-01  2.0734E-02  1.0974E-02  7.9570E-03
E (MeV)=  34.000 phase shift (rad.)=  1.4771E-02 -6.5504E-01 -6.6884E-01 -6.7773E-01
E (MeV)=  44.000 Amplitude=           5.3759E-01  1.6176E-01  2.0848E-01  2.1180E-01
E (MeV)=  44.000 phase shift (rad.)=  1.8357E-01  2.4807E-01  3.4413E-02 -7.6101E-02

total angular momentum= 30
number of basis functions per interval= 25
number of intervals=  4
channel radius=  14.0000
Number of energies:  2
Initial energy:   34.0000
Energy step:   10.0000
E (MeV)=  34.000 Amplitude=           9.9370E-01  2.0814E-02  1.1000E-02  7.9145E-03
E (MeV)=  34.000 phase shift (rad.)=  1.4832E-02 -6.5501E-01 -6.6880E-01 -6.7762E-01
E (MeV)=  44.000 Amplitude=           5.3757E-01  1.6177E-01  2.0848E-01  2.1177E-01
E (MeV)=  44.000 phase shift (rad.)=  1.8360E-01  2.4808E-01  3.4417E-02 -7.6024E-02

total angular momentum= 30
number of basis functions per interval= 50
number of intervals=  2
channel radius=  14.0000
Number of energies:  2
Initial energy:   34.0000
Energy step:   10.0000
E (MeV)=  34.000 Amplitude=           9.9370E-01  2.0814E-02  1.1000E-02  7.9145E-03
E (MeV)=  34.000 phase shift (rad.)=  1.4832E-02 -6.5501E-01 -6.6880E-01 -6.7762E-01
 Wave function at a fixed step size
 Channel  1
   2.000 -2.3419E-11 -2.4100E-11
   4.000 -8.4875E-05  9.9903E-05
   6.000  1.3439E-03 -2.6307E-04
   8.000  6.1952E-03  3.7707E-03
  10.000  1.6631E-02 -4.5625E-01
  12.000  5.1200E-02 -3.4580E+00
  14.000 -2.6606E-02  2.1608E+00
  16.000 -3.6187E-02  2.1613E+00
  18.000 -2.4200E-02  1.1868E+00
  20.000 -2.2727E-02  1.0954E+00
 Channel  2
   2.000 -1.5236E-10  2.5608E-10
   4.000  1.6747E-04  2.4738E-05
   6.000 -1.1584E-03  2.4194E-04
   8.000 -5.4272E-03 -7.5530E-03
  10.000 -1.4726E-02  2.4985E-02
  12.000 -3.2496E-02 -1.4804E-02
  14.000  1.2679E-02  2.6010E-02
  16.000  2.6244E-02 -3.9777E-03
  18.000  1.8569E-02 -1.7156E-02
  20.000  1.5656E-02 -1.8832E-02
 Channel  3
   2.000  1.3121E-11  1.2765E-11
   4.000  4.9824E-05 -6.0398E-05
   6.000 -8.3111E-04  8.6104E-05
   8.000 -3.5400E-03 -4.6614E-03
  10.000 -6.1700E-03  1.5577E-02
  12.000 -2.0150E-02  3.4619E-03
  14.000  1.5600E-02  2.2871E-03
  16.000  3.2947E-03 -1.3872E-02
  18.000 -6.5698E-03 -1.1795E-02
  20.000 -8.8601E-03 -9.5665E-03
 Channel  4
   2.000  1.3734E-12 -6.9826E-13
   4.000 -1.4966E-05 -4.1342E-05
   6.000 -9.2885E-04  1.5343E-05
   8.000 -3.5720E-03 -4.1435E-03
  10.000 -4.3643E-03  1.4468E-02
  12.000 -1.3143E-02  1.0031E-02
  14.000  8.5137E-03 -8.1312E-03
  16.000 -8.2890E-03 -6.3665E-03
  18.000 -9.5315E-03  2.3991E-03
  20.000 -7.7940E-03  5.3598E-03
E (MeV)=  44.000 Amplitude=           5.3757E-01  1.6177E-01  2.0848E-01  2.1177E-01
E (MeV)=  44.000 phase shift (rad.)=  1.8360E-01  2.4808E-01  3.4417E-02 -7.6024E-02
 Wave function at a fixed step size
 Channel  1
   2.000  5.1169E-10  2.2624E-09
   4.000  3.5023E-03 -1.8169E-03
   6.000 -1.5940E-02  3.4657E-02
   8.000 -4.2896E-03 -1.2585E-02
  10.000  3.9204E-01 -2.5708E+00
  12.000  6.1462E-01  9.4364E-02
  14.000  4.8419E-01  4.5397E-01
  16.000  6.2567E-01 -7.3889E-01
  18.000  2.6037E-01 -1.7221E+00
  20.000 -5.4054E-01  1.0385E-01
 Channel  2
   2.000  5.8255E-08 -8.2729E-09
   4.000 -9.1816E-03 -9.0830E-03
   6.000  5.6940E-02 -4.5175E-02
   8.000  4.0620E-01  1.4412E-02
  10.000  1.7297E-01 -2.1291E-01
  12.000 -1.5159E-01 -1.5453E-01
  14.000 -1.6344E-01 -1.1417E-01
  16.000 -6.0469E-02 -1.8073E-01
  18.000  1.3399E-01 -1.2785E-01
  20.000  1.3510E-01  1.2128E-01
 Channel  3
   2.000  1.2179E-09 -2.8657E-09
   4.000 -5.7318E-03 -1.0215E-03
   6.000  3.8482E-02 -6.1791E-03
   8.000  1.8361E-01 -1.8840E-01
  10.000 -1.1708E-01 -3.7177E-01
  12.000 -2.2134E-01  1.8160E-01
  14.000 -1.1709E-01  2.3285E-01
  16.000 -2.0146E-01  1.4434E-01
  18.000 -2.2135E-01 -9.3229E-02
  20.000  5.5096E-02 -2.2853E-01
 Channel  4
   2.000 -7.7840E-11 -2.7581E-10
   4.000 -3.8399E-03  9.1628E-04
   6.000  3.0656E-02  2.6539E-02
   8.000  1.3643E-01 -3.1429E-01
  10.000 -3.8360E-01 -2.3309E-01
  12.000  9.9274E-02  2.8378E-01
  14.000  2.2823E-01  1.4253E-01
  16.000  1.6361E-01  1.9467E-01
  18.000 -4.8941E-02  2.4076E-01
  20.000 -2.3968E-01  1.2736E-02
\end{verbatim}

\section*{Appendix D}
\begin{figure}[h]
	\centering
	\includegraphics*[width=.6\textwidth,clip=]{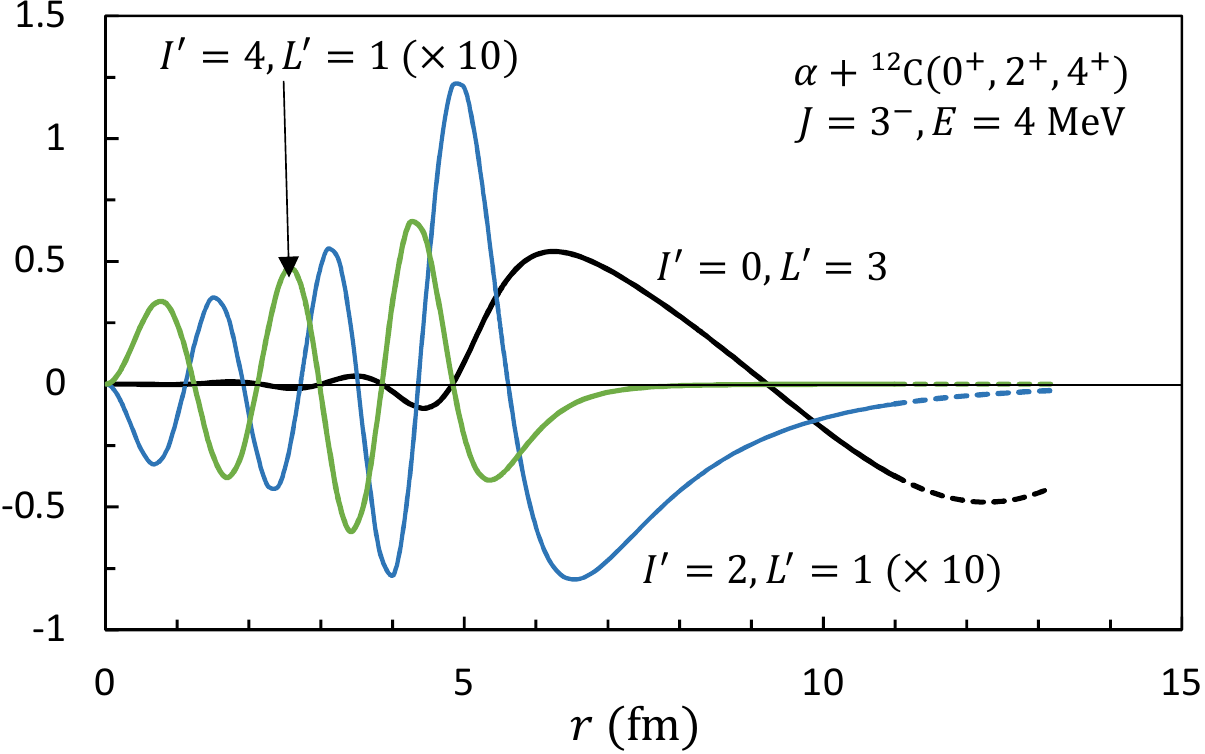}
	\caption{Real part of the $J=3^-$  \ac\ wave function at $\ecm=4$ MeV, with a channel 
		radius $a=11$ fm. We show the elastic ($I'=0,L'=3$) and two inelastic components 
		($I'=2,L'=1$ and $I'=4,L'=1$).
		The solid and dashed lines represent the internal and external wave functions, respectively.}
	\label{fig4}
\end{figure}

This appendix gives the output of example 4.
The wave function at $\ecm=4$ MeV is presented in Fig.\ \ref{fig4}.
\begin{verbatim}  
total angular momentum=  3
number of basis functions per interval= 25
number of intervals=  4
channel radius=   9.0000
Number of energies:  5
Initial energy:    4.0000
Energy step:    4.0000
E (MeV)=   4.000 Amplitude=           6.3596E-01
E (MeV)=   4.000 phase shift (rad.)= -9.4942E-02
E (MeV)=   8.000 Amplitude=           1.7408E-01  4.1658E-02  5.2298E-02  2.6680E-02
E (MeV)=   8.000 phase shift (rad.)= -1.0478E+00 -3.6134E-01 -9.3976E-02 -3.3158E-01
E (MeV)=  12.000 Amplitude=           7.7759E-02  3.8135E-02  3.0539E-02  7.3661E-02
E (MeV)=  12.000 phase shift (rad.)=  1.1344E+00  1.2203E+00 -8.6494E-01 -5.5400E-01
E (MeV)=  16.000 Amplitude=           4.2879E-02  2.7207E-02  2.9671E-02  3.4527E-02
                                      2.9332E-03  1.0489E-03  1.8585E-04  2.5162E-05
E (MeV)=  16.000 phase shift (rad.)=  3.3080E-01  4.2471E-01  1.2639E+00 -1.2662E+00
                                     -6.6359E-01 -9.7805E-01 -1.3598E+00  1.5139E+00
E (MeV)=  20.000 Amplitude=           2.8051E-02  1.9170E-02  1.9106E-02  2.0175E-02
                                      4.7665E-03  6.3974E-03  9.4889E-03  5.8624E-03
E (MeV)=  20.000 phase shift (rad.)= -4.1339E-01 -5.1731E-01  4.2013E-01  1.0489E+00
                                      7.7182E-01  1.3264E+00  1.4657E+00  1.2864E+00

Number of energies:  0
Initial energy:    0.0000
Energy step:    0.0000

total angular momentum=  3
number of basis functions per interval= 25
number of intervals=  4
channel radius=  10.0000
Number of energies:  5
Initial energy:    4.0000
Energy step:    4.0000
E (MeV)=   4.000 Amplitude=           6.3211E-01
E (MeV)=   4.000 phase shift (rad.)= -1.0498E-01
E (MeV)=   8.000 Amplitude=           1.8026E-01  4.3834E-02  5.2228E-02  2.7052E-02
E (MeV)=   8.000 phase shift (rad.)= -1.0365E+00 -3.7119E-02 -1.2120E-01 -3.7472E-01
E (MeV)=  12.000 Amplitude=           8.3944E-02  1.3480E-02  4.0382E-02  8.1238E-02
E (MeV)=  12.000 phase shift (rad.)=  1.1559E+00 -1.5076E+00 -8.3169E-01 -6.3383E-01
E (MeV)=  16.000 Amplitude=           4.3161E-02  2.7045E-02  2.9524E-02  3.4464E-02
                                      2.9445E-03  1.0500E-03  1.8528E-04  2.5063E-05
E (MeV)=  16.000 phase shift (rad.)=  3.3205E-01  4.2416E-01  1.2627E+00 -1.2686E+00
                                     -6.5988E-01 -9.7391E-01 -1.3555E+00  1.5185E+00
E (MeV)=  20.000 Amplitude=           2.8039E-02  1.9165E-02  1.9073E-02  2.0003E-02
                                      4.7594E-03  6.3858E-03  9.4702E-03  5.8594E-03
E (MeV)=  20.000 phase shift (rad.)= -4.1803E-01 -5.1340E-01  4.2415E-01  1.0520E+00
                                      7.7283E-01  1.3272E+00  1.4670E+00  1.2879E+00

Number of energies:  0
Initial energy:    0.0000
Energy step:    0.0000

total angular momentum=  3
number of basis functions per interval= 20
number of intervals=  4
channel radius=  11.0000
Number of energies:  5
Initial energy:    4.0000
Energy step:    4.0000
E (MeV)=   4.000 Amplitude=           6.1525E-01
E (MeV)=   4.000 phase shift (rad.)= -1.0040E-01
E (MeV)=   8.000 Amplitude=           1.8113E-01  6.6769E-02  4.8913E-02  2.4643E-02
E (MeV)=   8.000 phase shift (rad.)= -1.0734E+00 -1.5641E-01 -1.0245E-01 -4.0518E-01
E (MeV)=  12.000 Amplitude=           8.7310E-02  4.2829E-02  4.0388E-02  7.3597E-02
E (MeV)=  12.000 phase shift (rad.)=  1.0617E+00 -1.4261E+00 -9.3782E-01 -7.2888E-01
E (MeV)=  16.000 Amplitude=           4.3124E-02  2.7057E-02  2.9530E-02  3.4450E-02
                                      2.9433E-03  1.0489E-03  1.8491E-04  2.4997E-05
E (MeV)=  16.000 phase shift (rad.)=  3.3174E-01  4.2457E-01  1.2632E+00 -1.2680E+00
                                     -6.5924E-01 -9.7341E-01 -1.3553E+00  1.5188E+00
E (MeV)=  20.000 Amplitude=           2.8037E-02  1.9167E-02  1.9080E-02  2.0027E-02
                                      4.7582E-03  6.3842E-03  9.4671E-03  5.8584E-03
E (MeV)=  20.000 phase shift (rad.)= -4.1741E-01 -5.1378E-01  4.2378E-01  1.0520E+00
                                      7.7293E-01  1.3273E+00  1.4670E+00  1.2880E+00

Number of energies:  0
Initial energy:    0.0000
Energy step:    0.0000
\end{verbatim}

\section*{Appendix E}
\begin{figure}[h]
	\centering
	\includegraphics*[width=.6\textwidth,clip=]{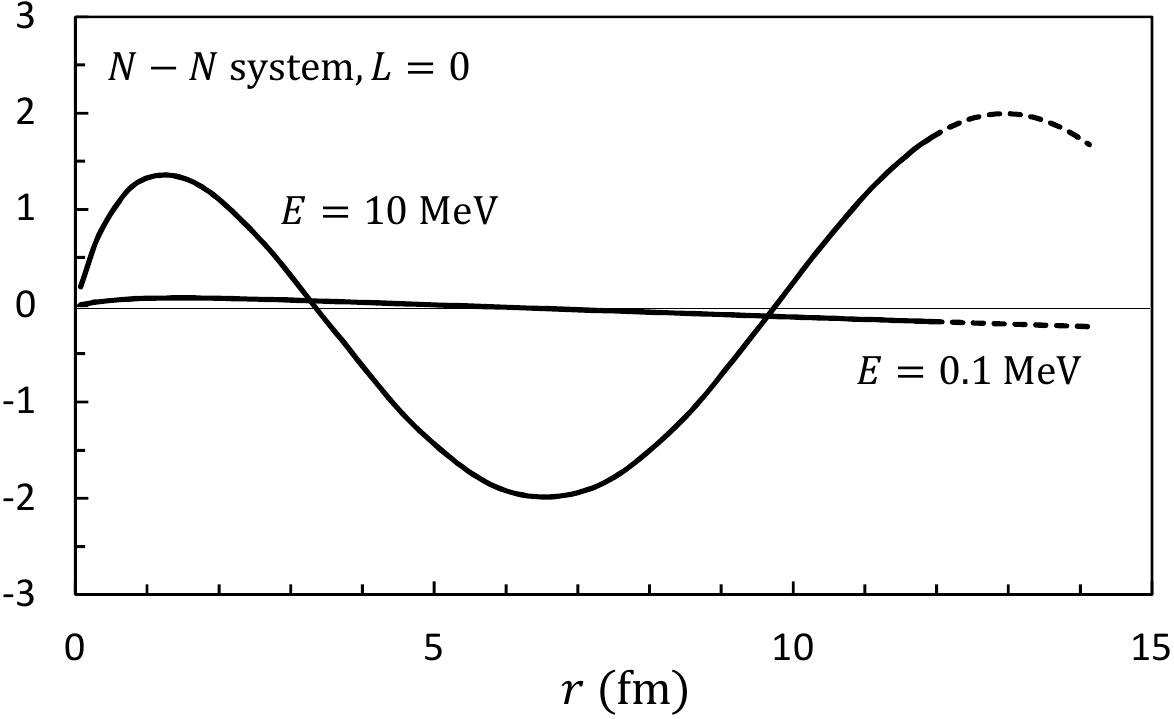}
	\caption{Real part of the $L=0$  nucleon-nucleon wave function with the non-local
		potential (\ref{eq_yam}) at two energies: $\ecm=0.1$ and 10 MeV. The channel 
		radius is $a=12$ fm.
		The solid and dashed lines represent the internal and external wave functions, respectively.}
	\label{fig5}
\end{figure}

This appendix gives the output of example 5.
The wave functions at $\ecm=0.1$ and 10 MeV are presented in Fig.\ \ref{fig5}.

\begin{verbatim} 

total angular momentum=  0
number of basis functions per interval= 10
number of intervals=  1
channel radius=   8.0000
Number of energies:  2
Initial energy:    0.1000
Energy step:    9.9000
E (MeV)=   0.100 phase shift (deg.)= -1.5077E+01  1.0000E+00
E (MeV)=  10.000 phase shift (deg.)=  8.5637E+01  1.0000E+00
Number of energies:  0
Initial energy:    0.0000
Energy step:    0.0000

total angular momentum=  0
number of basis functions per interval= 15
number of intervals=  1
channel radius=   8.0000
Number of energies:  2
Initial energy:    0.1000
Energy step:    9.9000
E (MeV)=   0.100 phase shift (deg.)= -1.5077E+01  1.0000E+00
E (MeV)=  10.000 phase shift (deg.)=  8.5637E+01  1.0000E+00
Number of energies:  0
Initial energy:    0.0000
Energy step:    0.0000

total angular momentum=  0
number of basis functions per interval= 15
number of intervals=  1
channel radius=  12.0000
Number of energies:  2
Initial energy:    0.1000
Energy step:    9.9000
E (MeV)=   0.100 phase shift (deg.)= -1.5079E+01  1.0000E+00
E (MeV)=  10.000 phase shift (deg.)=  8.5635E+01  1.0000E+00
Number of energies:  0
Initial energy:    0.0000
Energy step:    0.0000
 \end{verbatim}


\begin{thebibliography}{10}
	\expandafter\ifx\csname url\endcsname\relax
	\def\url#1{\texttt{#1}}\fi
	\expandafter\ifx\csname urlprefix\endcsname\relax\def\urlprefix{URL }\fi
	\expandafter\ifx\csname href\endcsname\relax
	\def\href#1#2{#2} \def\path#1{#1}\fi
	
	\bibitem{Ta65}
	T.~Tamura,  Rev. Mod. Phys. 37 (1965) 679.
	
	\bibitem{Sa83}
	G.~R. Satchler, Direct Nuclear Reactions, Oxford (1983).
	
	\bibitem{Th88}
	I.~J. Thompson, Comput. Phys. Rep. 7 (1988) 167.
	
	\bibitem{MTG98}
	L.~A. Morgan, J.~Tennyson, C.~J. Gillan, Comput. Phys. Commun. 114 (1998) 120.
	
	\bibitem{PN99}
	M.~Plummer, C.~J. Noble, J. Phys. B 32~(14) (1999) L345.
	
	\bibitem{Bu11}
	P.~Burke, R-Matrix Theory of Atomic Collisions. Application to Atomic,
	Molecular and Optical Processes, Vol.~61 of Springer Series on Atomic,
	Optical, and Plasma Physics, Springer, 2011.
	
	\bibitem{CH13}
	L.~F. Canto, M.~S. Hussein, Scattering Theory of Molecules, Atoms and Nuclei,
	World Scientific Publishing, Singapore, 2013.
	
	\bibitem{ZDF93}
	M.~V. Zhukov, B.~V. Danilin, D.~V. Fedorov, J.~M. Bang, I.~J. Thompson, J.~S.
	Vaagen, Phys. Rep. 231 (1993) 151.
	
	\bibitem{TND04}
	I.~J. Thompson, F.~M. Nunes, B.~V. Danilin, Comput. Phys. Commun. 161 (2004) 87.
	
	\bibitem{Ho77}
	H.~Horiuchi,  Prog.	Theor. Phys. Suppl. 62 (1977) 90.
	
	\bibitem{TMO07}
	M.~Theeten, H.~Matsumura, M.~Orabi, D.~Baye, P.~Descouvemont, Y.~Fujiwara,
	Y.~Suzuki, Phys. Rev. C 76 (2007) 054003.
	
	\bibitem{QN08}
	S.~Quaglioni, P.~Navr\'{a}til,  Phys. Rev. Lett. 101
	(2008) 092501.
	
	\bibitem{KNG01}
	H. Kamada, A. Nogga, W. Gl\"{o}ckle, E. Hiyama, M. Kamimura, K. Varga, Y. Suzuki, M. Viviani, A. Kievsky, S. Rosati, 
	J. Carlson, Steven C. Pieper, R. B. Wiringa, P. Navr\'{a}til, B. R. Barrett, N. Barnea, W. Leidemann, G. Orlandini,
	Phys. Rev. C 64 (2001) 044001
	
	\bibitem{HKK03}
	E.~Hiyama, Y.~Kino, M.~Kamimura, Prog. Part. Nucl.
	Phys. 51 (2003) 223.
		
	\bibitem{VKR06}
	M.~Viviani, A.~Kievsky, S.~Rosati, Phys.
	Rev. C 71 (2005) 024006.
	
	\bibitem{MBH13}
	J.~Mitroy, S.~Bubin, W.~Horiuchi, Y.~Suzuki, L.~Adamowicz, W.~Cencek,
	K.~Szalewicz, J.~Komasa, D.~Blume, K.~Varga, Rev. Mod. Phys. 85 (2013)
	693.
	
	\bibitem{VS95}
	K.~Varga, Y.~Suzuki,  Phys. Rev. C 52 (1995) 2885.
	
	\bibitem{Ra72}
	J.~Raynal, in "Computing as a Language of Physics", Trieste 1971, IAEA, Vienna
	(1972) p. 281.
	
	\bibitem{BP82b}
	W.~Baylis, S.~Peel,
		Comput. Phys. Commun. 25 (1982) 7.
	
	\bibitem{Th87}
	A. E. Thorlacius, E. D. Cooper, J. Comput. Phys. 72 (1987) 70.
	
	\bibitem{Wi46}
	E.~P. Wigner, Phys. Rev. 70 (1946) 606.
	
	\bibitem{TDE00}
	I.~J. Thompson, B.~V. Danilin, V.~D. Efros, J.~S. Vaagen, J.~M. Bang, M.~V.
	Zhukov, Phys. Rev. C 61 (2000) 024318.
	
	\bibitem{DTB06}
	P.~Descouvemont, E.~M. Tursunov, D.~Baye, Nucl. Phys. A 765 (2006) 370.
	
	\bibitem{BCD09}
	D.~Baye, P.~Capel, P.~Descouvemont, Y.~Suzuki,  Phys. Rev. C 79
	(2009) 024607.
	
	\bibitem{LT58}
	A.~M. Lane, R.~G. Thomas, Rev. Mod. Phys. 30 (1958) 257.
	
	\bibitem{DB10}
	P.~Descouvemont, D.~Baye, Rep. Prog. Phys. 73 (2010)
	036301.
	
	\bibitem{Ba15}
	D.~Baye, Phys. Rep. 565 (2015) 1.
	
	\bibitem{HSV98}
	M.~Hesse, J.~M. Sparenberg, F.~Van~Raemdonck, D.~Baye, Nucl. Phys. A 640 (1998)
	37.
	
	\bibitem{HRB02}
	M.~Hesse, J.~Roland, D.~Baye, Nucl. Phys. A 709 (2002) 184.
	
	\bibitem{BBM82}
	K.~L. Baluja, P.~G. Burke, L.~A. Morgan, Comput. Phys. Commun. 27
	(1982) 299.
	
	\bibitem{Mo84}
	L.~A. Morgan, Comput. Phys. Commun. 31 (1984) 419.
	
	\bibitem{BHT68}
	F.~C. Barker, H.~J. Hay, P.~B. Treacy, Aust. J. Phys. 21 (1968) 239.
	
	\bibitem{Th10}
	I.~J. Thompson, NIST Handbook of Mathematical Functions, Cambridge University
	Press, 2010, p. 741.
	
	\bibitem{Ol10}
	A.~B.~Olde Daalhuis, NIST Handbook of Mathematical Functions, Cambridge
	University Press, 2010, p. 321.
	
	\bibitem{Bl57}
	C.~Bloch, Nucl. Phys. 4 (1957) 503.
	
	\bibitem{BRT83}
	R.~F. Barrett, B.~A. Robson, W.~Tobocman, Rev. Mod. Phys. 55 (1983) 155.
	
	\bibitem{Bu67}
	P.~J.~A. Buttle, Phys. Rev. 160 (1967) 719.
	
	\bibitem{LW76}
	J.~C. Light, R.~B. Walker, J. Chem. Phys. 65 (1976)
	4272.
	
	\bibitem{BD83}
	D.~Baye, P.~Descouvemont, Nucl. Phys. A 407 (1983) 77.
	
	\bibitem{BHS98}
	D.~Baye, M.~Hesse, J.-M. Sparenberg, M.~Vincke, J. Phys. B 31 (1998) 3439.
	
	\bibitem{DBD10}
	T.~Druet, D.~Baye, P.~Descouvemont, J.-M. Sparenberg, Nucl. Phys. A 845 (2010) 88.
	
	\bibitem{BGS02}
	D.~Baye, J.~Goldbeter, J.-M. Sparenberg, Phys. Rev. A 65 (2002) 052710.
	
	\bibitem{BHV02}
	D.~Baye, M.~Hesse, M.~Vincke, Phys. Rev. E 65 (2002) 026701.
	
	\bibitem{Sl11}
	V.~Szalay, T.~Szidarovszky, G.~Czak\'{o}, A.~G. Cs\'{a}sz\'{a}r, Journal of Mathematical Chemistry 50 (2011) 636.
	
	\bibitem{DD12b}
	T.~Druet, P.~Descouvemont,  Eur. Phys. J. A 48 (2012) 147.
	
	\bibitem{RMP80}
	M.~Rhoades-Brown, M.~H. Macfarlane, S.~C. Pieper, Phys. Rev. C
	21 (1980) 2417.
	
	\bibitem{La92} Lapack library, http://www.netlib.org/lapack.
	
	\bibitem{AS72}
	M.~Abramowitz, I.~A. Stegun, Handbook of Mathematical Functions, Dover, London,
	1972.
	
	\bibitem{ZJ96}
	S.~Zhang, J.~M. Jin, Computation of Special Functions, Wiley-Interscience,
	1996, source available at http://jin.ece.illinois.edu/routines/routines.html.
	
	\bibitem{Ba84b}
	A.~Barnett, Comput. Phys. Commun. 35 (1984) C812.
\bibitem{PFT86}
W.~H. Press, B.~P. Flannery, S.~A. Teukolsky, W.~T. Vettering, Numerical
Recipes, Cambridge University Press (1986).	
	\bibitem{GSW70}
	G.~Goldring, M.~Samuel, B.~A. Watson, M.~C. Bertin, S.~L. Tabor, Phys. Lett. B 32 (1970) 465.
	
	\bibitem{Re68}
	R.~V. Reid, Ann. Phys. 50 (1968) 411.
	
	\bibitem{RMP80b}
	M.~Rhoades-Brown, M.~H. Macfarlane, S.~C. Pieper, Phys. Rev. C 21 (1980) 2436.
	
	\bibitem{RNT01}
	S.~Raman, C.~W. Nestor, P.~Tikkanen, At. Data Nucl. Data Tables 78 (2001) 1.
	
	\bibitem{Ya65}
	Y.~Yamaguchi, Phys. Rev. 95 (1954) 1628.
	
\end{thebibliography}

\end{document}